\documentclass{iopjournal}

\usepackage{amssymb} 
\usepackage{amsmath} 
\usepackage{color}

\newcommand{\changed}[1]{{#1}}

\begin{document}

\articletype{Paper} 

\title{Bayesian optimization of stellarator alpha-particle confinement using data-informed parameter spaces and dimensionality reduction}

\author{
Matt Landreman$^{1,*}$\orcid{0000-0002-7233-577X},
Michael Czekanski$^2$\orcid{0009-0005-2520-3415},
Andrew Giuliani$^3$\orcid{0000-0002-4388-2782},
Byoungchan Jang$^1$\orcid{0000-0003-1097-352X},
Rory Conlin$^1$\orcid{0000-0001-8366-2111}
}

\affil{$^1$Institute for Research in Electronics and Applied Physics, University of Maryland, College Park, MD, USA}

\affil{$^2$Department of Statistics and Data Science, Cornell University, Ithaca, NY, USA}

\affil{$^3$Flatiron Institute, New York, NY, USA}

\affil{$^*$Author to whom any correspondence should be addressed.}

\email{mattland@umd.edu}


\newcommand{\mcnote}[1]{\textcolor{blue}{(Michael: #1)}}

\begin{abstract}
Modern stellarators are typically designed by optimizing the shape of the plasma boundary surface, with the parameters taken to be Fourier amplitudes.
Many promising optimization algorithms such as Bayesian methods require bound constraints on the parameters and are most efficient when each parameter is scaled similarly to the others.
With the typical Fourier parameterization, it is unclear how to set these bounds: wide constraints lead to self-intersecting boundaries and frequent failures of the MHD equilibrium calculation, while tight bound constraints limit expressiveness.
To address these issues, here we propose two new parameter spaces for stellarator optimization.
Both begin with a dataset of existing stellarator boundaries.
In the first approach, a quantile transformation is applied to each Fourier degree of freedom, mapping the data distribution to a uniform distribution on the unit interval.
In the second approach, principal component analysis (PCA) is applied to points on the boundaries, followed by a quantile transformation.
For both approaches, the transformed variables become the degrees of freedom, naturally bounded to [0, 1].
The PCA method has the additional benefit of dimensionality reduction, with high expressiveness for a small number of parameters.
The methods are demonstrated via Bayesian optimization for good alpha-particle confinement with guiding-center tracing inside the optimization loop, using asynchronous parallelization.
These optimizations yield stellarator configurations with excellent \changed{collisionless} fast-particle confinement in fields that can be far from quasisymmetric or quasi-isodynamic.
\end{abstract}


\section{Introduction}

In modern stellarator design, the first step is to optimize the shape of the plasma boundary to achieve adequate confinement and stability.
Such an optimization is necessary because otherwise the neoclassical radial transport and losses of energetic particles would usually be too large, and the design may not be stable to pressure-driven magnetohydrodynamic (MHD) modes.
This optimization is typically done using a Fourier description of the plasma boundary, taking the Fourier amplitudes to be the degrees of freedom (i.e. design variables).
However, this parameter space of Fourier amplitudes has several significant disadvantages.
Optimization algorithms are most effective when the objective has a similar scale length in each dimension of the input space,
but high-order Fourier modes generally have much smaller amplitudes than low-order modes \cite{jang2026exponential}, and the rate of decrease with mode number is unclear.
Moreover, many promising optimization algorithms, particularly global methods, require bound constraints (also known as box constraints), and with a Fourier description it is unclear how to set these bounds.
If the constraints are set to allow too wide a range, the boundary surface may become self-intersecting and unphysical, such that an MHD equilibrium cannot be computed and the objective cannot even be evaluated.
On the other hand, if the bound constraints are set to too narrow an interval, the domain does not include a diverse range of geometries, and the optimal geometry may be missed.

To address these issues, in this paper we present two new methods to define the parameter space for stellarator shape optimization.
Both methods provide natural scaling of the parameters, while balancing the competing considerations of maximizing expressiveness and minimizing the fraction of the domain that is self-intersecting.
The two approaches are based on first collecting a set of known stellarator boundary shapes.
These data effectively provide a prior probability distribution on the space of boundary shapes.
In the first method, which retains a typical Fourier representation of the boundary, a quantile transformation of the data is used to define a transformed version of each Fourier amplitude that varies between 0 and 1.
In the second method, principal component analysis (PCA) is first applied to the boundary data, and then a quantile transformation is applied to each principal component's amplitude.
In both methods, weights can be introduced in case certain geometries are over-represented or under-represented in the data.
The two approaches also provide natural bound constraints on each parameter, for use with optimization algorithms that require these constraints.
Both transformations are piecewise-linear and hence compatible with gradient-based optimization methods.

Another shortcoming of the traditional Fourier parameter space is that the relatively high dimensionality required makes it hard to include high-fidelity calculations in the objective.
Local gradient-based optimization methods have been quite successful in finding stellarator geometries with favorable physics properties \cite{landreman2022magnetic, goodman2023constructing, giuliani2024direct,dudt2024magnetic,liu2026optimizingstellaratorshiddensymmetry}, with objective functions that include theory-based surrogates for confinement such as the deviation from quasisymmetry or omnigenity.
However, these theory-based surrogates are only partially correlated with the factors that truly matter.
In particular, measures of quasisymmetry or omnigenity are imperfectly correlated with energetic particle confinement \cite{bader2021modeling, paul2022energetic,wiedman2024coil,hegna2025infinity}, and proxies for turbulent transport are only partially correlated with actual simulations of turbulence \cite{landreman2025does}.
It would therefore be advantageous to directly optimize the factors that we truly care about.
By optimizing the quantities that directly matter rather than loosely-correlated surrogates, it may be possible to discover new stellarator geometries that meet the necessary core physics criteria, giving more freedom to satisfy other objectives and constraints such as those related to engineering or cost.
However, both energetic particle confinement (typically computed with Monte Carlo methods) and turbulence involve chaos, resulting in objective functions that are not smooth functions of the plasma shape, likely necessitating gradient-free optimization methods.
Gradient-free algorithms scale poorly with the number of dimensions, so it would be desirable to lower the number of parameters required to describe stellarator boundaries.
The PCA method here provides this advantage.

To demonstrate an application of the new parameter spaces, we will perform Bayesian optimization (BO) to find geometries with good alpha-particle confinement.
BO is a state-of-the-art method for derivative-free global optimization that takes full advantage of the information in all past evaluations of the objective function.
Due to this sample-efficiency and BO's ability to handle noisy objectives, it should be well suited for optimization of Monte Carlo guiding-center calculations.
In this paper, direct tracing of guiding centers is performed inside the optimization loop using an efficient GPU implementation \cite{czekanski2026catapult}.
Using a novel objective function based on the time it takes to reach a threshold value of alpha losses, rather than the losses at a fixed time, computation is preferentially used at the end of the optimization to refine the best configurations.
This approach, resembling a multi-fidelity method, ensures that minimal computation is allocated to poorly performing geometries.
An asynchronous algorithm is employed so the objective function can be evaluated at different points in parameter space concurrently, accounting for the wide variations in time to evaluate the objective, while keeping all the GPUs busy.
We present five new optimized stellarator configurations generated using these methods, including ones with a constraint for Mercier stability.
All these stellarators have good confinement of energetic particles and low thermal neoclassical transport, and all show significant departures from ideal quasisymmetric, omnigenous, and quasi-isodynamic field patterns.
These findings add to the mounting evidence \cite{bindel2023direct,velasco2024piecewise} that quasisymmetry and omnigenity are overly constraining conditions, at least for the size and field scale considered here.

Several prior publications have discussed issues related to the topics of this paper.
In \cite{jang2026exponential}, an exponential scaling of Fourier modes for stellarator boundaries was shown to improve efficiency of derivative-based optimization.
While this exponential scaling partially addresses the aforementioned problem of parameter-space anisotropy, the methods here account more fully for the prior probability distribution of each parameter, while being robust to outliers and providing a principled approach to bound constraints.
Principal component analysis of stellarator boundaries was used previously in \cite{giuliani2024comprehensive}.
There it was used to discover that quasisymmetric stellarators of given rotational transform and number of field periods tend to lie along curves  in parameter space and exist in a small number of clusters.
Cadena et al \cite{cadena2025constellaration} used PCA along with a Gaussian mixture model and Markov-chain Monte Carlo to generate new boundary shapes from data.
Wei et al used PCA and an autoencoder to find low-dimensional representations of quasi-helically symmetric equilibria \cite{wei2026low}.
The PCA method in our work differs from these earlier studies in its combination with quantile transformation and application to optimization.
Ali et al \cite{ali2025poster} devised a real-space surface representation in which bound constraints are straightforward, and applied a Bayesian algorithm to optimize for quasisymmetry.
Bayesian optimization has also been applied previously to stellarator coils \cite{glas2022global,giuliani2024direct}.
Optimizations using different algorithms with guiding-center tracing inside the optimization loop have been described in \cite{gori2001alpha,ku2008physics,bindel2023direct}.
Overall, the new contributions in our paper include both the novel surface parameterizations and their demonstration in Bayesian optimization of guiding-center tracing inside the objective function.

In the following section, the two new parameter spaces for stellarator optimization are defined in more detail.
Then in section \ref{sec:example}, a specific dataset is used to demonstrate the methods, and the parameter spaces derived from this dataset are explored.
As an application, in section \ref{sec:optimization} we present optimizations of alpha-particle confinement, with direct tracing of guiding-center trajectories inside the optimization loop.
We discuss the results and conclude in section \ref{sec:conclusions}.


\section{Methods}
\label{sec:methods}


\subsection{Quantile transformation}
\label{sec:quantile_transformation}

\begin{figure}
 \centering
        \includegraphics[width=\textwidth]{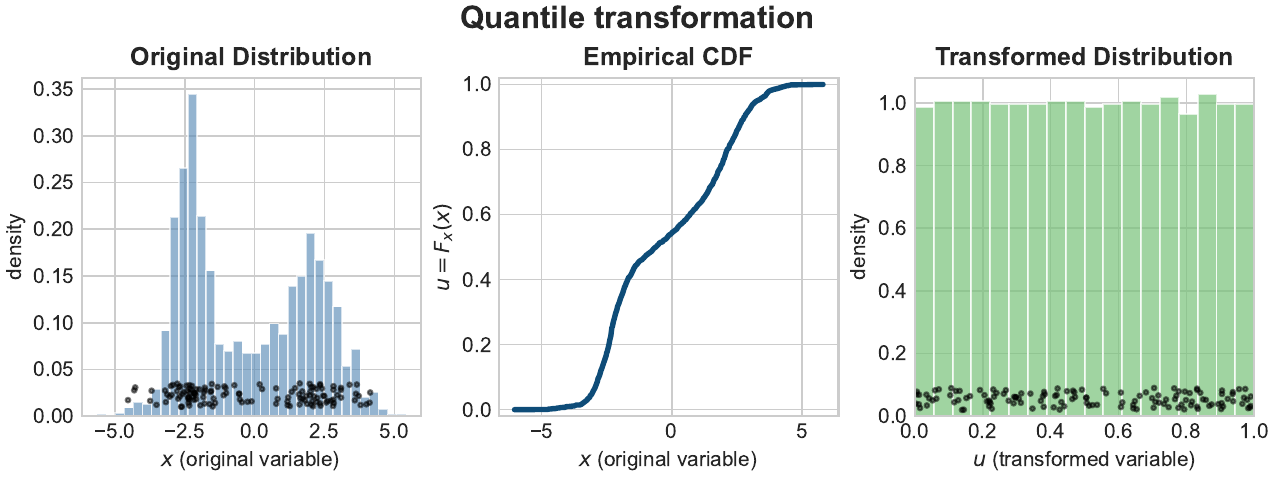}
 \caption{A quantile transformation is a piecewise-linear function that maps an arbitrary data distribution to a uniform distribution on $[0,1]$.
 In the left and right panels, black dots show typical data points with a random vertical location so the points can be distinguished.
 }
\label{fig:quantile_transformation}
\end{figure}

A key component of both parameter spaces in our approach is a quantile transformation.
A quantile transformation is a map $F_x$ that transforms an arbitrary data distribution to a uniform distribution on the unit interval $[0,1]$.
Instead of using one of the original degrees of freedom $x$ in an optimization, we use the transformed variable $u = F_x(x)$ as the degree of
freedom.
Quantile transformation is illustrated in figure \ref{fig:quantile_transformation}.

The map $F_x$ is the empirical cumulative distribution function (CDF), estimated from the training samples of that parameter.
The function $F_x(x)$ is monotonic and is chosen to be piecewise-linear.
Given the value of the new degree of freedom $u$, the original variable $x$ is recovered by applying the inverse map $x = F_x^{-1}(u)$.
The inverse map is then obtained straightforwardly by linear interpolation between empirical quantiles.
(While a CDF could alternatively be defined from data via a piecewise-constant function, this choice is not useful in our optimization context since it would not be invertible.)
Restating these points in practical terms, the map from $x$ to $u$ is \texttt{interp(x\_data, u\_data)} where \texttt{interp} is a 1D interpolant based on the arguments, \texttt{x\_data} is the array of sorted data values, and \texttt{u\_data = linspace(0, 1, len(x\_data))} is a linearly spaced array of the \texttt{len(x\_data)} values going from 0 to 1.
Similarly, the inverse map is \texttt{interp(u\_data, x\_data)}.
This quantile transformation is available in scikit-learn \cite{scikit-learn} through the \texttt{QuantileTransformer} class with \texttt{output\_distribution='uniform'}.

For an optimization problem in more than one variable, an independent quantile transformation is applied to each parameter.
In doing so we neglect correlations between the parameters, but in practice this minimal approach is \changed{adequate enough to obtain several of the optimized configurations that will be shown in section \ref{sec:results}}.
The transformed parameters all have comparable scales to each other, $O(1)$, as is preferable for many optimization algorithms.
Natural bound constraints on each parameter $u$ are $0 \le u \le 1$, corresponding to the original parameter $x$ lying within the range of the data.
If a prior probability for $u$ is taken to be a uniform distribution on $[0,1]$, as would be assumed at the start of Bayesian optimization, this corresponds to a prior on $x$ matching the data distribution, which is quite reasonable.
Note that no assumption is made as to the data distribution, so it can be far from a normal distribution, and can be multi-modal.
Also, since the transformation is piecewise-linear, it is differentiable almost everywhere, and so is compatible with gradient-based optimization.

The quantile transformation is preferable to a simpler linear or affine transformation for several reasons.
Quantile transformation is robust to outliers in the data, as they get compressed to a small fraction of the space.
For example, given 101 data points with one outlier on the high end, only the top 1\% of the domain ($u \in [0.99,1]$) is affected by the outlier.
At the same time, outliers are not ignored, and the extreme values of any parameter in the data are achieved when $u=0$ or $u=1$.
For each parameter, the quantile transformation incorporates information about the entire distribution of values in the data, not merely the range or standard deviation of values, without any assumption that the values are distributed normally.
In contrast, with a linear or affine transformation, if bound constraints are chosen to include the entire range of data, then an outlier causes the bound constraints to expand significantly, compressing the majority of the data to a small subset of the search space.
If, however, outliers are dropped when setting bound constraints with a linear or affine transformation, then not all values in the data are achievable.

A common situation is that the training data contains many samples from one class and few samples of another class.
For example, thanks to databases such as QUASR \cite{giuliani2024direct,giuliani2024comprehensive}, a dataset of plasma shapes can include a very large number of quasisymmetric configurations, but for noteworthy real stellarators such as W7-X or LHD, each may have only one or a few variants in the dataset.
Therefore, a weighted version of the quantile transformation can be established so the distribution is not overwhelmed by the class of configurations that are more numerous.
Each configuration in the sample data is assigned a weight by the user.
If all configurations are given a weight $w$ except for one configuration that has a weight $wp$ for some number $p$, the weighted transformation is equivalent to the unweighted transformation in which that configuration is repeated in the data $p$ times.
As with the standard quantile transformation, each transformed parameter $u$ lies in $[0, 1]$.


\subsection{Fourier parameter space}
\label{sec:fourier}

Now we discuss the first of the two parameter spaces for optimization.
In this method, a separate quantile transformation is applied to each Fourier mode describing the plasma boundary.
There are three Fourier descriptions commonly used for stellarators, and the method here is applicable to any of the three.
All the Fourier descriptions refer to cylindrical coordinates $(R,\phi,Z)$, where $\phi$ is the toroidal angle, and to an arbitrary poloidal angle $\theta$.
One common Fourier description is 
\begin{equation}
R(\theta,\phi) = \sum_{m,n}R_{m,n}\cos(m\theta-n_{fp} n\phi),
\;\;\;
Z(\theta,\phi) = \sum_{m,n}Z_{m,n}\sin(m\theta-n_{fp} n\phi),
\label{eq:Fourier}
\end{equation}
where $m$ and $n$ are integers, $n_{fp}$ is the number of field periods, and $(R_{m,n}, Z_{m,n})$ are the parameters.
(We will assume stellarator symmetry for simplicity, though the methods of this paper can be applied straightforwardly in the non-stellarator-symmetric case.)
A second Fourier description is Garabedian's, defined using complex numbers:
\begin{equation}
       R(\theta,\phi) + i Z(\theta,\phi) = e^{i\theta} \sum_{m,n} \Delta_{m,n} e^{-im \theta + i n_{fp} n \phi},
\end{equation}
where $\Delta_{m,n}$ are the parameters \cite{anderson1994stellarator,garabedian1995reduction}.
A third way stellarator boundaries have been described using Fourier series is\cite{dudt2020desc}

\begin{equation}
\begin{array}{l}
R(\theta,\phi) = \sum_{m,n} \bar{R}_{m,n} \mathcal{G}_{m,n}(\theta,\phi), \\
Z(\theta,\phi) = \sum_{m,n} \bar{Z}_{m,n} \mathcal{G}_{m,n}(\theta,\phi),
\end{array}
\end{equation}
where
\begin{equation}
\mathcal{G}_{m,n}(\theta,\phi)=
\left\{
\begin{array}{ll}
\cos\left(|m|\theta\right)\cos\left(n_{fp} |n|\phi\right), & m,n\geq 0, \\
\cos\left(|m|\theta\right)\sin\left(n_{fp} |n|\phi\right), & m\geq 0,\, n<0, \\
\sin\left(|m|\theta\right)\cos\left(n_{fp} |n|\phi\right), & m<0,\, n\geq 0, \\
\sin\left(|m|\theta\right)\sin\left(n_{fp} |n|\phi\right), & m,n<0.
\end{array}
\right.
\end{equation}
In this third approach the parameters are $\bar{R}_{m,n}$ and $\bar{Z}_{m,n}$.
The parameters in the three approaches are linear combinations of each other, as can be shown with trigonometric identities.

We may now define the first parameter space for optimization.
Let $x$ denote any of the Fourier parameters; an independent quantile transformation is applied to each $x$, and the transformed variables $u$ become the new degrees of freedom.
This approach can be applied to any of the three Fourier representations above.

The method has an adjustable level of conservatism in the space of possible boundary shapes, which the user can control by choosing bound constraints that are a subset or superset of $[0,1]$.
For example, by choosing each parameter to lie in $[0.1, 0.9]$, the most extreme 10\% of the parameter values at the high end and at the low end from the data are avoided.
This makes it more likely that the equilibrium calculation will converge well, with the trade-off that it is no longer possible to realize the most extreme boundary shapes.
Conversely, if the interpolant for the inverse quantile transformation is extended to allow extrapolation, then bound constraints outside the unit interval such as [-0.1, 1.1] can be allowed.
This choice allows the optimizer to explore values of the Fourier parameters beyond the range of the data.
\changed{Such extrapolation was not used for any results in this paper.}


\subsection{Major radius, minor radius, and aspect ratio}

To complete the specification of the data-informed Fourier parameter space, some additional discussion is necessary regarding the major and minor radius and how they can be controlled when optimizing in this space.
Typically in optimizations, the user wishes to specify the major radius, minor radius, or both.
One definition of minor radius $a$ is given by $\pi a^2=\bar{S}$ where $\bar{S}=(2\pi)^{-1}\int d\phi \; S$ is the toroidal average of the area $S(\phi)$ of the outer surface’s cross-section in the $RZ$ plane.
One definition of major radius $R_0$ then follows from $(2\pi R_0)(\pi a^2)=V$ where $V$ is the enclosed volume.
These definitions are used for instance by the codes VMEC and DESC.
Alternative non-equivalent definitions of major and minor radius are the Garabedian amplitudes $\Delta_{1,0}$ and $\Delta_{0,0}$ respectively.
Two non-equivalent possible definitions of  aspect ratio are $A = R_0 / a$ or $A_G = \Delta_{1,0} / \Delta_{0,0}$.

Generally the input data will contain configurations that vary in major and minor radius, which may also differ from the major and minor radius of the desired configuration.
For this reason we take the following steps.
First we scale all configurations in the input data to the same minor radius.
Then for given optimization parameters $\{ u_j \}$, after applying the transformation to Fourier amplitudes $\{ x_j \}$, the amplitudes are multiplied by the desired minor radius.
In the most straightforward approach, Garabedian amplitudes are chosen for the $\{ x_j \}$.
The two parameters $\Delta_{1,0}$ and $\Delta_{0,0}$ are not included in the parameter space, and instead they are set directly to their desired values.
This helps to further reduce the dimensionality of the optimization problem, and there is no need to include an objective or constraint to control the aspect ratio or plasma size, simplifying the formulation of the optimization problem.

A potential downside of fixing $\Delta_{1,0}$ and $\Delta_{0,0}$ is that these values can deviate somewhat from the more commonly used $R_0$ and $a$ for strongly shaped boundaries (and so $A_G$ can deviate from the more commonly used $A$).
If a user prefers to fix $R_0$ and $a$, this can be achieved as follows.
The data are normalized by the Garabedian minor radius $\Delta_{0,0}$ to define the quantile transformations.
At each function evaluation in the optimization, the inverse quantile transformations are applied to compute temporary $\Delta_{m,n}$ amplitudes.
Fixing $\Delta_{0,0}=1$, the Garabedian major radius $\Delta_{1,0}$ is varied in a 1D root solve so that the surface overall matches the target $A$.
(The cost of this root solve is negligible compared to that of evaluating a typical objective.)
Then $a$ is computed, and all the $\Delta_{m,n}$ amplitudes (including major and minor radius) are multiplied by $a_{\mathrm{target}} / a$ so the resulting surface has minor radius $a_{\mathrm{target}}$.


\subsection{Parameter space from principal component analysis}

\begin{figure}
 \centering
        \includegraphics[width=4in]{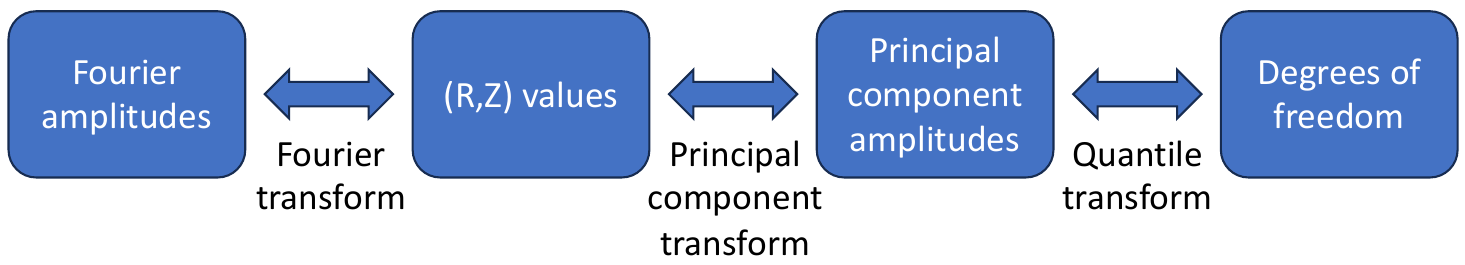}
 \caption{Transformation pipeline between boundary Fourier amplitudes and PCA degrees of freedom.}
\label{fig:pipeline}
\end{figure}

With the first data-informed parameter space now fully specified, we now define the second parameter space, which is based on PCA.
The first step is to assemble a data matrix $X$ based on $(R,Z)$ values on each boundary in the supplied data.
To do this, each configuration in the dataset is scaled to the same minor radius $a$, a uniform tensor-product grid in the angles $(\theta,n_{fp}\phi)$ is fixed, and $R$ and $Z$ are evaluated at these points for each boundary.
The $n_{fp}\phi$ grid covers half a field period for every configuration, regardless of $n_{fp}$.
The average major radius of each boundary is subtracted from $R$ to obtain $R' = R - R_0$.
The resulting 2D arrays are flattened so that for each configuration, the $R'$ and $Z$ values each form a row vector, then these row vectors are concatenated to make a single vector of twice the length.
The vectors from each configuration are stacked vertically to form $X$.
The mean and first $N$ principal components are computed.
From the amplitudes of the principal components over the data, a quantile transformation is defined, as in section \ref{sec:quantile_transformation}.
Thus, a pipeline of multiple transformations has been fit to the data, as shown in figure \ref{fig:pipeline}.
Given a new point in the optimization parameter space, the standard boundary Fourier amplitudes are computed by applying the pipeline of transformations from right to left.
First  the inverse quantile transform is applied, then  the resulting PC amplitudes are transformed to $(R',Z)$ values.
They are scaled up to the desired minor radius, and the desired major radius is added to $R'$.
Finally the $(R,Z)$ values are Fourier transformed.
Alternatively, the PCA could be performed directly on Fourier amplitudes instead of on real-space coordinates.
The results in this case are observed to be extremely similar, as expected from Parseval's theorem equating the norms of functions and their Fourier coefficients.
Attractive features of the real-space approach are that it is clear that all real-space coordinates can be given equal weight in the PCA, and surfaces with different parameterizations can be combined in the same PCA.

The PCA surface parameterization described above results in an approximate but not exact match to the minor and major radius definitions $a$ and $R_0$.
If desired, an exact match can be achieved with an additional step.
Instead of adding $R_0$ to $R'$, a 1D root solve is performed to find the value to be added to $R'$ such that the resulting surface has exactly the target aspect ratio $A$.
Then, the boundary is uniformly scaled to match the target major and minor radius.

As discussed in section \ref{sec:quantile_transformation}, a likely scenario is that samples from one class may be over-represented in the data, while there are few samples from another class.
In order for each class to contribute more equitably to the PCs, a weighted PCA can be used.
The weight may differ for each row of $X$ but it is the same for all columns, so we do not need the most general form of weighted PCA.
In our case, weights can be accounted for by first computing a weighted mean of each column, then subtracting this weighted mean to obtain a centered data matrix $\bar{X}$.
Next, a weighted covariance matrix is formed, $\bar{X}^T W \bar{X}$, where $W$ is a diagonal matrix $W_{jj} = (w_j / (\sum_i w_i))(n/(n-1))$ containing the weights $w_j$ for the $n$ samples, $j=1, \ldots, n$. 
The eigenvectors of this covariance matrix are the weighted principal components.
Our implementation of the weighted PCA and weighted quantile transformations is available at \url{github.com/byoungj/weightedpca}.

\changed{
The Fourier parameter space described in section \ref{sec:fourier} represents a prior distribution on boundary shapes that crudely approximates the data by matching the marginal distribution of each Fourier mode.
The PCA parameter space provides an improved prior that also accounts for linear correlations, though it still will not fully represent the data distribution if there are nonlinear dependencies.
}


\subsection{What range of data to include?}

For both the Fourier method and PCA method, one question that arises is whether the boundary shapes in the dataset should have the same aspect ratio and/or number of field periods as the new configuration one wishes to optimize.
On the one hand, constant-$\phi$ cross-sections that yield good properties at one aspect ratio and $n_{fp}$ may not necessarily result in good physics properties if the major radius or $n_{fp}$ is changed.
By this reasoning, there may be value in limiting the dataset to include only boundaries of the same $n_{fp}$ and similar aspect ratios to the values that will be used in optimization.
However, this approach significantly restricts the number of configurations in the dataset.
In principle there is no reason that configurations from multiple values of $n_{fp}$ and aspect ratio cannot be included in the data.
Note that the major and minor radius and $n_{fp}$ are set independently of the data in both the Fourier and PCA methods.
\changed{Pooling configurations with differing $n_{fp}$ is not unreasonable since cross-sections of typical optimized stellarators are similar functions of $n_{fp}\phi$, especially for $n_{fp} \ge 3$, with tall bean-shaped sections at $\phi=0$ and short wide sections at $\phi=\pi/n_{fp}$.}
A boundary shape that is free of self-intersections at one value of $n_{fp}$ will also be self-intersection-free at any other $n_{fp}$; it will also be free of self-intersections if the major radius is altered to change the aspect ratio unless the new aspect ratio is extremely low.
Or, as a compromise, one could choose to exclude from the data only configurations with $n_{fp}=1$, or only those with $n_{fp} \le 2$, since the cross-sections for those numbers of field periods are most different from cross-sections at $n_{fp} \ge 3$.
For the results that follow, we combine configurations of multiple $n_{fp}$ values and a wide range of aspect ratios in the data, unless stated otherwise.
Users of our methods may wish to try the inclusive, exclusive, and compromise approaches to see which works better for a specific application.
Regardless of this choice, the data and weights define only the optimization coordinates; any resulting boundary can be expressed in standard Fourier amplitudes independent of the dataset used.


\section{Numerical Examples}
\label{sec:example}


\subsection{Usable fraction, diversity, and baseline}
\label{sec:baseline}

For any attempt at Bayesian or other global optimization of a stellarator geometry, an important metric is the fraction $f_{\mathrm{us}}$ of the hypercube design space (within the bound constraints) for which the MHD equilibrium code converges.
We will call $f_{\mathrm{us}}$ the ``usable fraction'' of the design space.
If $f_{\mathrm{us}}$ is too low, the initial sampling of the space by the optimization algorithm will not generate enough meaningful values of the objective function to guide the search, and the optimization will fail.
The usable fraction depends on the equilibrium code and the resolution parameters used; we define success if \texttt{vmec++} \cite{schilling2025numerics} converges to a normalized force residual of $10^{-11}$ with 51 radius surfaces, and poloidal and toroidal Fourier resolution of 8, allowing a maximum of 5000 iterations.
Higher $f_{\mathrm{us}}$ is obtained with a larger number of iterations or higher force residual.
The usable fraction also depends on the choice of $n_{fp}$ and aspect ratio; for calculations here we fix $n_{fp}=4$ and $A=6$.
The normalized plasma pressure $\beta$ also affects the usable fraction; we use a pressure profile and $B$ scaling that will be described in detail in section \ref{sec:profiles}, resulting in volume-averaged $\beta$ of 1.5-3\% depending on the geometry, unless noted otherwise.
To compute $f_{\mathrm{us}}$ with any given choice of parameters, we randomly sample $\gtrsim 4000$ points uniformly in the design space; the fraction of these points for which \texttt{vmec++} converges to the requested force residual then gives $f_{\mathrm{us}}$.

\begin{figure}
 \centering
\includegraphics[width=0.43\textwidth]{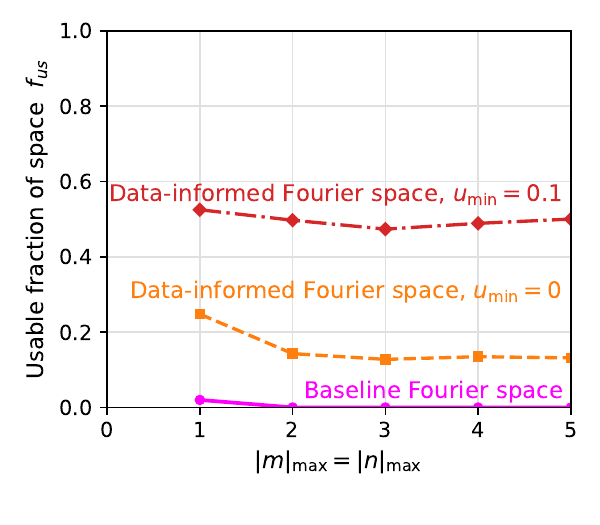}
\includegraphics[width=0.53\textwidth]{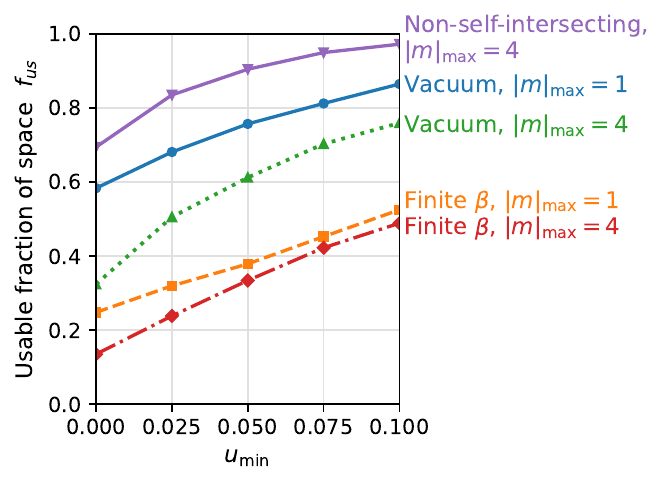}
 \caption{Fraction $f_{\mathrm{us}}$ of the baseline and data-informed Fourier parameter spaces for which the MHD equilibrium converges to the required tolerance.
Here, $|m|_{\max}$ and $|n|_{\max}$ refer to the maximum $|m|$ and $|n|$ values in eq (\ref{eq:Fourier}).
 For the baseline, the width of the bound constraints was optimized to maximize $f_{\mathrm{us}}$.
 For the data-informed space, each shape parameter is randomly sampled uniformly on $[u_{\min}, 1-u_{\min}]$.
Values in the left panel are for finite $\beta$, and values on the right are for the data-informed space.
The right panel also shows the fraction of the space for which the boundaries do not self-intersect.
 }
\label{fig:usable_fraction_Fourier}
\end{figure}

For a baseline, figure \ref{fig:usable_fraction_Fourier}.a shows $f_{\mathrm{us}}$ for a standard Fourier representation.
For this parameter space we keep Fourier mode amplitudes $m$ and $n$ in eq (\ref{eq:Fourier}) up to maximum values $|m|_{\max}=|n|_{\max}$, corresponding to Garabedian amplitudes $\Delta_{m',n}$ with $|n| \le |n|_{\max}$ and $|m'-1| \le |m|_{\max}$.
The Garabedian major and minor radius are fixed, while the other Fourier amplitudes are allowed to vary within symmetric bound constraints, $|\Delta_{m,n}| \le x_{m,n}^{\max}$.
Keeping the Garabedian minor radius fixed helps to reduce self-intersections.
Motivated by \cite{jang2026exponential}, the bound constraints are tighter for higher mode numbers according to $x_{m,n}^{\max} = x_0 \exp(-\sqrt{m^2+n^2})$ where $x_0$ is chosen by the user.
At each value of $|m|_{\max}=|n|_{\max}$, the width of the bound constraints $x_0$ is scanned to find the value giving the highest $f_{\mathrm{us}}$.
These acheived values of $f_{\mathrm{us}}$ are extremely small: 2\% for $|m|_{\max}=|n|_{\max}=1$, and 0.04\% for $|m|_{\max}=|n|_{\max}=2$.
For tight bound constraints (small values of $x_0$), the geometry is not far from an axisymmetric torus and too weakly shaped to provide much rotational transform, giving an equilibrium $\beta$ limit that is too small to support the prescribed pressure profile, resulting in failure of the equilbrium code to converge.
For wide bound constraints (high values of $x_0$) and $|m|_{\max}=|n|_{\max}>1$, larger $\iota$ can be achieved, making it possible to support the given pressure, but it also becomes likely for the boundary to self-intersect.
To describe interesting stellarator shapes, $|m|_{\max}=|n|_{\max}\geq 2$ is typically needed.
The very small values of $f_{\mathrm{us}}$ for the baseline Fourier representation make it unlikely that Bayesian or global optimization methods can succeed with this baseline space, motivating the new data-informed parameter spaces.

\begin{figure}
 \centering
\includegraphics[width=3in]{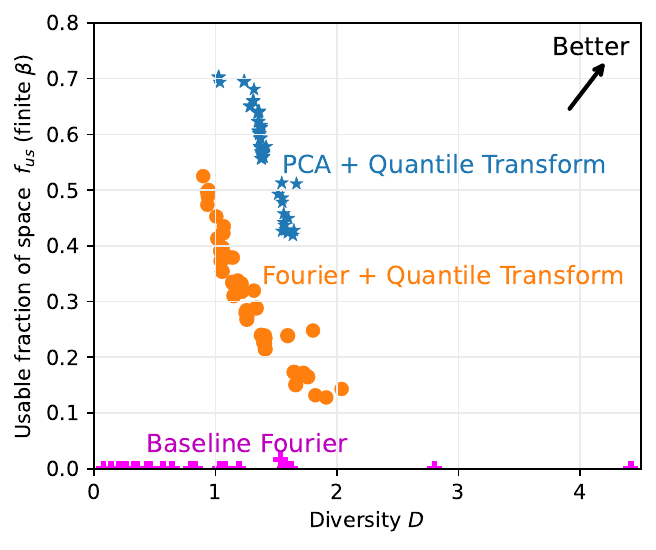}
 \caption{Pareto fronts reflecting the trade-off between diversity of possible shapes vs. the fraction of the space where the MHD equilibrium converges, as defined in section \ref{sec:baseline}.
 Each point represents a different choice of parameter space. 
 Within each of the three shape parameterizations (colors), the different points correspond to different choices of $|m|_{\max} = |n|_{\max}$, $u_{\min}$, $x_0$, number of principal components, and/or which values of $n_{fp}$ are included in the input data.
 }
\label{fig:Pareto}
\end{figure}

Besides $f_{\mathrm{us}}$, another important property of a bound-constrained parameter space for stellarator boundaries is the range of possible shapes that can be represented.
This property could be quantified in a variety of ways.
Here we define one such measure which is convenient and fast to compute, which we call the diversity $D$: the expected absolute difference in cylindrical coordinates between two randomly sampled surfaces from the space.
To compute $D$, two shapes are sampled randomly by drawing the parameters uniformly within the bounds.
The cylindrical coordinates are computed across the surfaces on a uniform tensor product grid in $(\theta,\phi)$.
The pointwise absolute difference $(|R_1 - R_2| + |Z_1 - Z_2|)/a$ (where subscripts refer to the two surfaces) is evaluated, and a mean over $\theta$ and $\phi$ is computed.
This calculation is repeated and averaged over many pairs of points in the parameter space to find the expected value.
\changed{As a caveat, $D$ depends not only on the range of expressible shapes but also on the sampling density, so it is better interpreted as a measure of typical variation rather than the maximum variation possible.}
The diversity for the baseline Fourier space is displayed in figure \ref{fig:Pareto}.
The different points correspond to different choices of $x_0$ and $|m|_{\max} = |n|_{\max}$.
For the baseline space, $D$ can range from small to large values depending on these choices, extending to very large values off the right of the plot for large $x_0$.

\subsection{Data preparation}
\label{sec:data_prep}

To illustrate the data-informed parameter spaces we assemble a diverse data set of existing stellarator boundaries.
The collection is drawn from three sources: the set of well-known configurations from Kappel et al \cite{kappel2024magnetic}, the QUASR database of quasisymmetric configurations \cite{giuliani2024direct, giuliani2024comprehensive}, and the ConStellaration database of quasi-isodynamic configurations \cite{cadena2025constellaration}.
The first set includes experimental stellarator configurations such as W7-X, HSX, LHD, TJ-II, NCSX, CTH, CNT, ATF, etc., as well as configurations from recent publications on optimization \cite{landreman2022magnetic, goodman2023constructing, sanchez2023quasi}.
For QUASR and ConStellaration, a subset of the best configurations is selected; details of this selection are given in \ref{sec:data_prep_details}.
The total number of boundary shapes used for the results that follow is 270.
In each of the three groups of equilibria, the weight of each sample for PCA and the quantile transformation is set to the inverse of the number of samples in the group, so each of the three groups contributes equally.

For both the Fourier and PCA parameter spaces, it is important that all boundary shapes in the input data be oriented consistently, as discussed in section 5.1 of \cite{giuliani2024comprehensive}.
Therefore, if necessary, each configuration is rotated or flipped, or the orientation of $\theta$ is reversed, to meet several conditions: $\theta$ should increase counter-clockwise in a constant-$\phi$ plane, 
$R(\theta=0, \phi=0)$ should exceed $R(\theta=\pi,\phi=0)$, $R(\theta=\pi,\phi=0)$ should exceed $R(\theta=\pi,\phi=\pi/n_{fp})$, and the $(m=0,n=1)$ mode of $Z$ should be positive.
Also, for all of the surfaces in the input data, we apply spectral condensation \cite{hirshman1985optimized}.
This is done since there is degeneracy in the poloidal angle, and for greatest economy of the principal components, they should ideally represent true changes in the boundary shape rather than mere reparameterizations.
\changed{The condensation algorithm used is described in \ref{sec:condensation}, which differs from the method in \cite{hirshman1985optimized}.}
Finally, all boundary shapes are scaled to a minor radius of 1.0.
For the PCA method, $R$ and $Z$ are evaluated on a tensor-product grid of $19\times 20$ points in $n_{fp}\phi$ and $\theta$ over one half field period.


\subsection{Data-informed Fourier method}

Figure \ref{fig:usable_fraction_Fourier}.a shows that $f_{\mathrm{us}}$ is far higher for the data-informed Fourier space than for the baseline Fourier space.
If the bound constraints are tightened from $[0, \,1]$ to $[u_{\min}, \, 1-u_{\min}]$ to exclude the most strongly shaped boundaries, $f_{\mathrm{us}}$ is increased.
When $u_{\min}=0.1$ the usable fraction at finite $\beta$ reaches 50\% even with a large number of Fourier modes.
This value of $f_{\mathrm{us}}$ is large enough for successful Bayesian optimization, as we will see below.
\changed{It may be advantageous to use a $u_{\min}$ comparable to this value only for high-order Fourier modes while retaining $u_{\min}=0$ for low-order Fourier modes, but for simplicity we use the same $u_{\min}$ for all modes in the work here.}
Figure \ref{fig:usable_fraction_Fourier} also shows that $f_{\mathrm{us}}$ decreases slightly with increasing $|m|_{max}$ and $|n|_{max}$, which is expected since as these parameters increase the boundaries become more strongly shaped.
Figure \ref{fig:usable_fraction_Fourier}.b shows that $f_{\mathrm{us}}$ is higher for vacuum than for finite $\beta$.
This trend is expected since it is typically harder to achieve a given force residual with increasing $\beta$, and for some boundary geometries, the given pressure profile may be beyond the equilibrium $\beta$ limit.
For comparison, figure \ref{fig:usable_fraction_Fourier}.b also shows the fraction of the space for which the surfaces are free of self-intersections (purple curve at top), irrespective of any equilibrium calculation.
This condition is less restrictive than convergence of an equilibrium code so the fraction is somewhat higher.

\begin{figure}
 \centering
        \includegraphics[width=\textwidth]{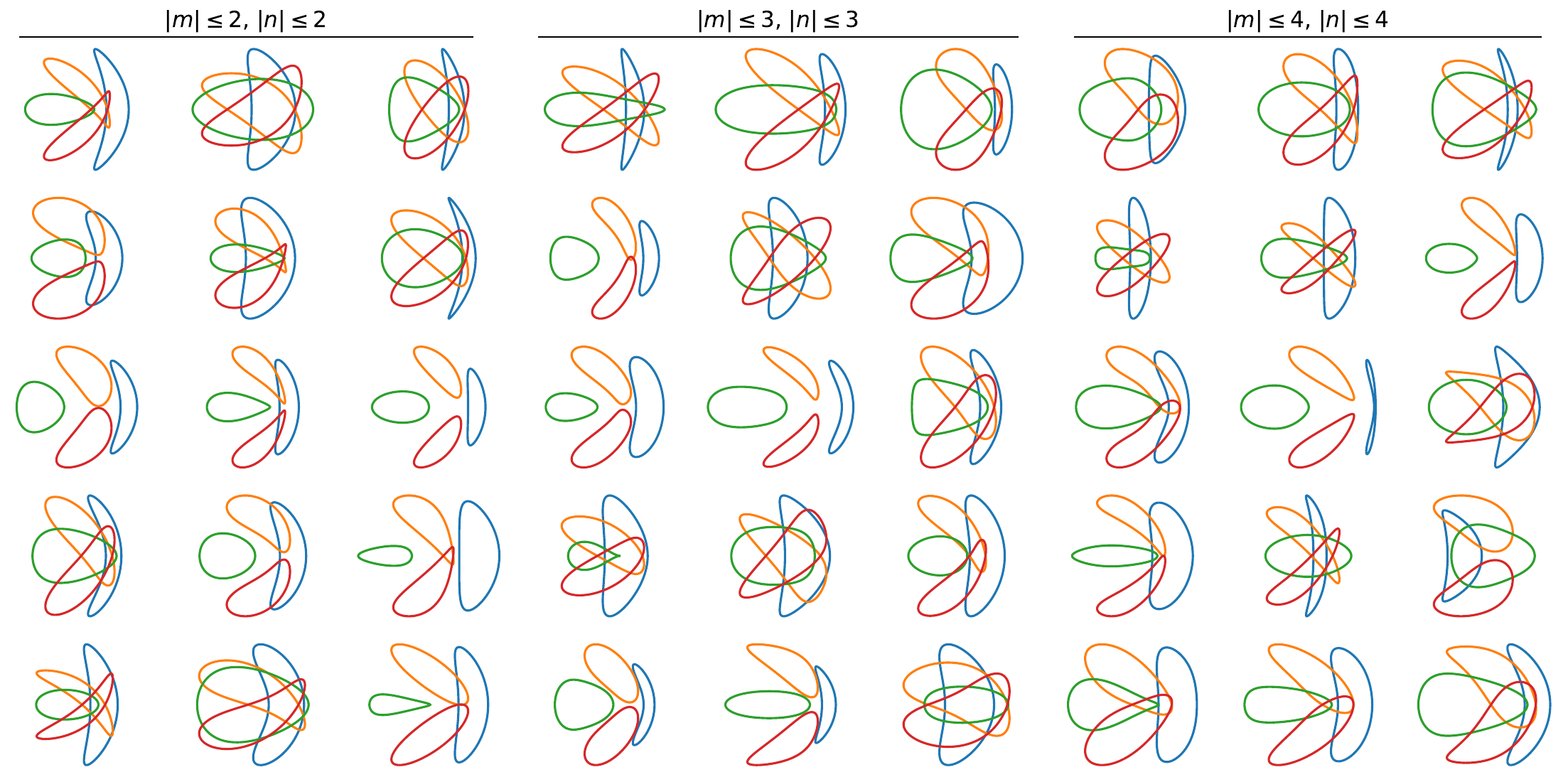}
 \caption{Cross-sections for stellarator shapes using the data-informed Fourier method.
 The degrees of freedom are randomly sampled evenly over the optimization space.
 A wide range of plausible strongly-shaped boundaries can be represented with few self-intersections.
 }
\label{fig:xsections_Fourier}
\end{figure}

Figure \ref{fig:xsections_Fourier} shows typical cross-sections for boundary shapes generated using the data-informed Fourier method.
To convey the distribution of typical shapes, the degrees of freedom are randomly sampled evenly over the parameter space.
In the figure, the domain of each design variable is $[0.1, 0.9]$, which is sufficiently conservative to eliminate all self-intersections in the boundaries shown.
Nonetheless it can be seen that a wide range of strongly shaped boundaries can be represented.
The figure shows results for maximum values of the poloidal and toroidal mode numbers in eq (\ref{eq:Fourier}) of 2, 3, or 4, corresponding respectively to spaces of 23, 47, and 79 dimensions.

In figure \ref{fig:Pareto}, the trade-off can be seen between $f_{\mathrm{us}}$ vs diversity $D$ for the data-informed Fourier space.
Points with higher $D$ and lower $f_{\mathrm{us}}$ are obtained when $u_{\min}$ is small and $|m|_{\max}=|n|_{\max}$ is large, ranging over the same values as in figure \ref{fig:usable_fraction_Fourier}.


\subsection{Principal component analysis method}

\begin{figure}
 \centering
        \includegraphics[width=3in]{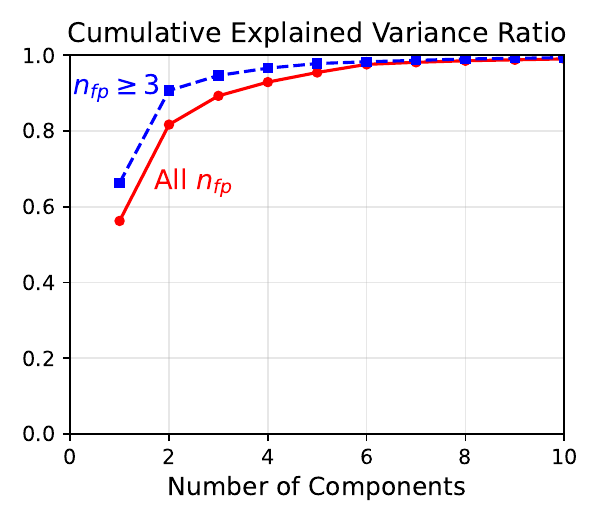}
 \caption{Most of the variance in the boundary shapes in the dataset is explained by a small number of principal components. 
 }
\label{fig:explained_variance}
\end{figure}

Figure \ref{fig:explained_variance} displays the variance among the boundary shapes in the dataset explained by the first few principal components.
It can be seen that the first few components explain much of the variance.
If the dataset is limited to configurations with $n_{fp} \ge 3$, fewer components are needed to reach the same level of explained variance.
This is because the collection includes boundaries with $n_{fp}=1$ and 2 that are quite different from the others, e.g. the configuration from \cite{jorge2022single}.

\begin{figure}
 \centering
        \includegraphics[width=\textwidth]{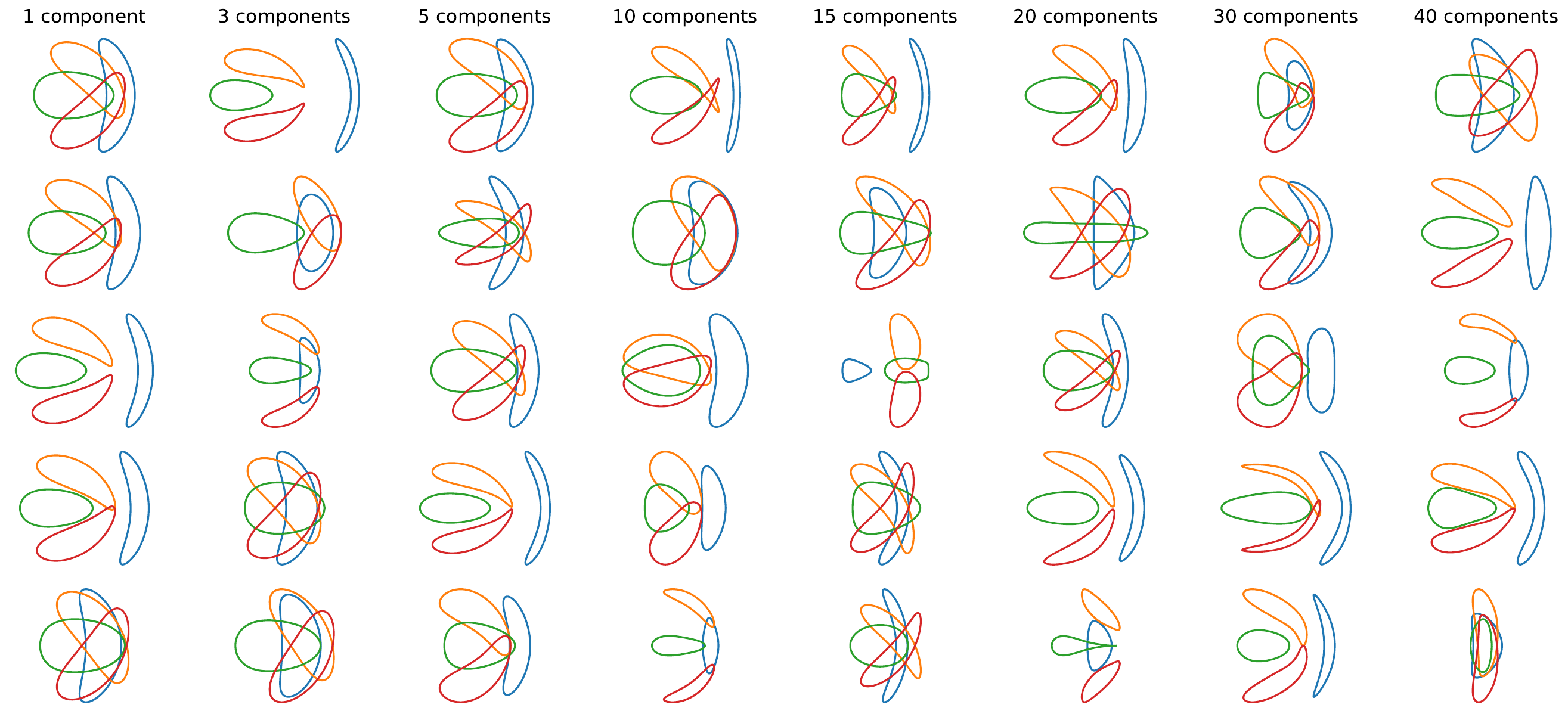}
 \caption{Cross-sections for stellarator shapes using the PCA method.
 The degrees of freedom are randomly sampled over the unit hypercube.
 A wide range of plausible strongly-shaped boundaries can be represented with few self-intersections.
 }
\label{fig:xsections_PCA}
\end{figure}

Figure \ref{fig:xsections_PCA} shows typical cross-sections for boundary shapes generated using the PCA method.
To convey the distribution of typical shapes, the degrees of freedom are randomly sampled within their range (i.e. the unit hypercube).
It can be seen in the figure that a wide range of strongly shaped boundaries can be represented.
At the same time, there are no apparent self-intersections.
From the variation within the leftmost column, in which only one principal component is retained, it can be seen that strong shaping is present even with a single component, and this first variable primarily controls the helical excursion of the magnetic axis.

\begin{figure}
 \centering
        \includegraphics[width=3.5in]{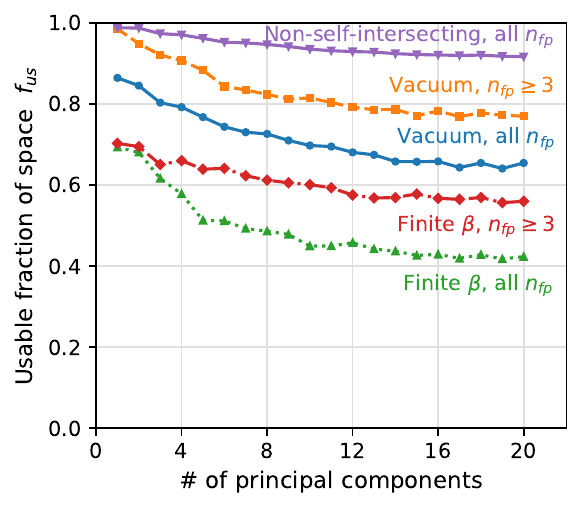}
 \caption{Fraction $f_{\mathrm{us}}$ of the PCA parameter space for which the MHD equilibrium converges to the required tolerance.
 Each parameter is randomly sampled uniformly on $[0,1]$.
 Also shown is the fraction of the space for which the surfaces are free of self-intersection (purple).
 }
\label{fig:usable_fraction_PCA}
\end{figure}

In figure \ref{fig:usable_fraction_PCA}, the usable fraction $f_{\mathrm{us}}$ is plotted for the PCA spaces, for several choices of parameters.
For these calculations, the full range $[0,1]$ is used for each parameter.
As with the Fourier approach, we find that $f_{\mathrm{us}}$ is higher for vacuum than for finite $\beta$.
We also observe that $f_{\mathrm{us}}$ is higher if boundaries from $n_{fp} \leq 2$ geometries are excluded from the dataset.
This trend makes sense because the $n_{fp}=1$ and $n_{fp}=2$ geometries have quite different shapes than the $n_{fp} \geq 3$ geometries.
There is a general decrease in $f_{\mathrm{us}}$ with increasing number of PCA components, which is expected since as more components are included the boundaries can become more strongly shaped.
The figure also shows the fraction of the space for which the surfaces do not self-intersect, which is higher than $f_{\mathrm{us}}$ due to the less stringent condition and which remains $> 91\%$.
Together, figures \ref{fig:xsections_PCA} and \ref{fig:usable_fraction_PCA} show that the PCA method simultaneously allows reasonably high $f_{\mathrm{us}}$ at the same time as a rich variety of highly shaped boundaries.

This same conclusion can be seen in figure \ref{fig:Pareto}.
The PCA data pushes the Pareto front between $f_{\mathrm{us}}$ vs diversity $D$ further in the desired direction compared to the other parameter spaces, even beyond the data-informed Fourier space.
For the PCA points on the plot, higher $D$ and lower $f_{\mathrm{us}}$ are obtained when the number of principal components is increased, and when the $n_{fp}=1$ and 2 configurations are included in the data.


\section{Application to optimization of alpha-particle confinement}
\label{sec:optimization}


\subsection{Profiles and scale}
\label{sec:profiles}

Now we demonstrate an application of the Fourier and PCA parameter spaces to optimization of fusion-produced alpha particle confinement.
This confinement depends on the scale of the device (size and magnetic field strength) and the initial distribution of the alpha particles, so we now discuss the choices for these factors.

\begin{figure}
 \centering
        \includegraphics[width=\textwidth]{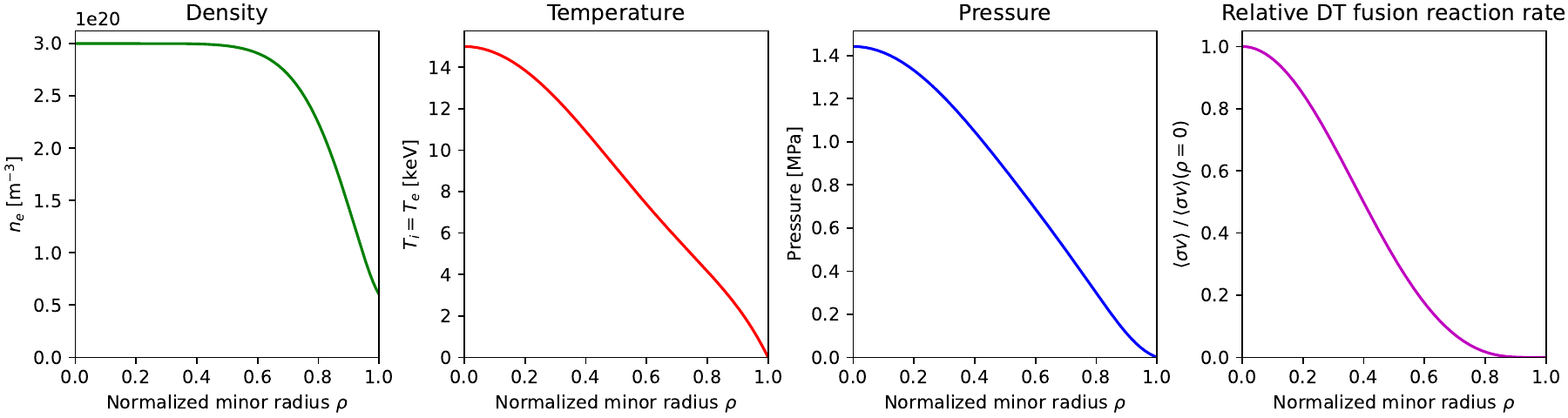}
 \caption{Profiles from section \ref{sec:profiles} used for the MHD equilibria and the alpha particle birth distribution.}
\label{fig:profiles}
\end{figure}

The birth distribution of fusion-produced alpha particles depends on the profiles of the main-species density and temperature.
We choose the electron density to be
\begin{equation}
n_e(\rho) = (3\times 10^{20}\;m^{-3})(1 - 2\rho^8+1.2\rho^{12}),
\end{equation}
with the density of deuterium and tritium each taken to be half of $n_e$.
The temperature profiles are chosen to be
\begin{equation}
T_i(\rho)=T_e(\rho) = (15 \; keV)(1 - 2\rho^2 + 2\rho^4 -\rho^6).
\end{equation}
These profiles are chosen to roughly reflect the profiles in recent reactor studies for high-field stellarators \cite{alonso2022physics, lion2025stellaris,hegna2025infinity, guttenfelder2025predictions}, while also being even polynomials of moderate order for simplicity.
The pressure profile $p = (T_e + T_i) n_e$ is then
\begin{equation}
p(\rho) = (1.44\times 10^6\; Pa)(1 -2\rho^2 + 2\rho^4 -\rho^6 -2\rho^8 + 4\rho^{10} -2.8\rho^{12} -0.4\rho^{14}+ 2.4\rho^{16} -1.2\rho^{18}).
\label{eq:pressure_profile}
\end{equation}
These profiles along with the ensuing DT fusion reaction rate are shown in figure \ref{fig:profiles}.
The pressure profile is used when computing MHD equilibria, and alpha particles are initialized volumetrically using the local fusion reaction rate and volume Jacobian using rejection sampling.

For the magnetic field scaling, we adjust the toroidal flux so the maximum field in the plasma is 12 T.
This value is chosen to be comparable to the values in several recent reactor designs based on high-temperature superconducting magnets: 12.7 T in \cite{lion2025stellaris}, 12.3 T in \cite{hegna2025infinity, schmitt2025magnetohydrodynamic}, and 11.3 T in \cite{landreman2026efficient}.
We opt to fix the maximum field rather than the mean because it is the maximum that is constrained by engineering factors, not the mean.
Note that fixing the maximum field rather than the mean provides a natural mechanism for the optimization to favor low mirror ratio: at high mirror ratio and fixed $B_{max}$, the minimum $B$ in the plasma would be small, leading to wider guiding-center orbits and hence higher losses.
To fix $B_{max}$, each equilibrium calculation is run twice: first with a guess for the toroidal flux, then with a toroidal flux scaled by $(12 \;T)/B_{max,actual}$ where $B_{max,actual}$ is the maximum $B$ in the plasma from the first calculation.
This represents one step of a fixed-point (Picard) iteration; more iterations could be done to more accurately achieve $B_{max} = 12\;T$ but we find that one is sufficient.
The initial guess for the toroidal flux is chosen to be high, since with a fixed pressure profile, low values of the toroidal flux may erroneously lead to non-convergence of the equilibrium code.
With $B_{max}= 12$ T and the pressure profile above, the volume-average $\beta$ is typically 1.5\%--3\% depending on the details of the plasma geometry.

For the size scaling, we note that high aspect ratio reactor designs such as Stellaris (\cite{lion2025stellaris}, $a=1.3$ m) and the Infinity Two design in \cite{hegna2025infinity} ($a=1.3$ m) tend to have lower minor radius than low aspect ratio reactors such as Helios (\cite{swanson2025overview}, $a=1.8$ m) and ARIES-CS (\cite{najmabadi2008aries}, $a=1.7$ m).
Performing a power series regression on these four configurations gives a scaling $a = (3.1 \mbox{m}) / A^{0.38}$ where $A$ is the aspect ratio.
	This scaling correctly fits the minor radius of these four configurations to two significant digits.

For all equilibria in this work we take the profile of toroidal current to be zero for simplicity.
In future work one could either optimize for small bootstrap current so this profile assumption was self-consistent, or else evaluate the current profile using a calculation of bootstrap current.


\subsection{Alpha particle confinement calculations}

To evaluate the objective for any particular boundary shape parameters, an MHD equilibrium is first computed using the code \texttt{vmec++}  \cite{schilling2025numerics}.
Then, guiding center trajectories of 3.5 MeV alpha particles are followed in Boozer coordinates without collisions using the CATAPULT GPU implementation \cite{czekanski2026catapult} in the package \texttt{firm3d}.
For each plasma configuration, 25,000 particles are followed, to approximately fill each Nvidia A100 GPU.
Tracing more than 25,000 particles would significantly increase the computation time because then some threads on the GPU would need to trace multiple particles in sequence.
\changed{For comparison, typical numbers of particles used in prior publications are 5,000-20,000 \cite{bader2019stellarator,paul2022energetic,bindel2023direct,goodman2024quasi}.
Including collisions might increase losses, as particles in confined trajectories could then scatter into loss cones, so including collisions will be an important element of future work.
}
A particle is considered lost as soon as it exits the equilibrium boundary.

\begin{figure}
 \centering
        \includegraphics[width=4.2in]{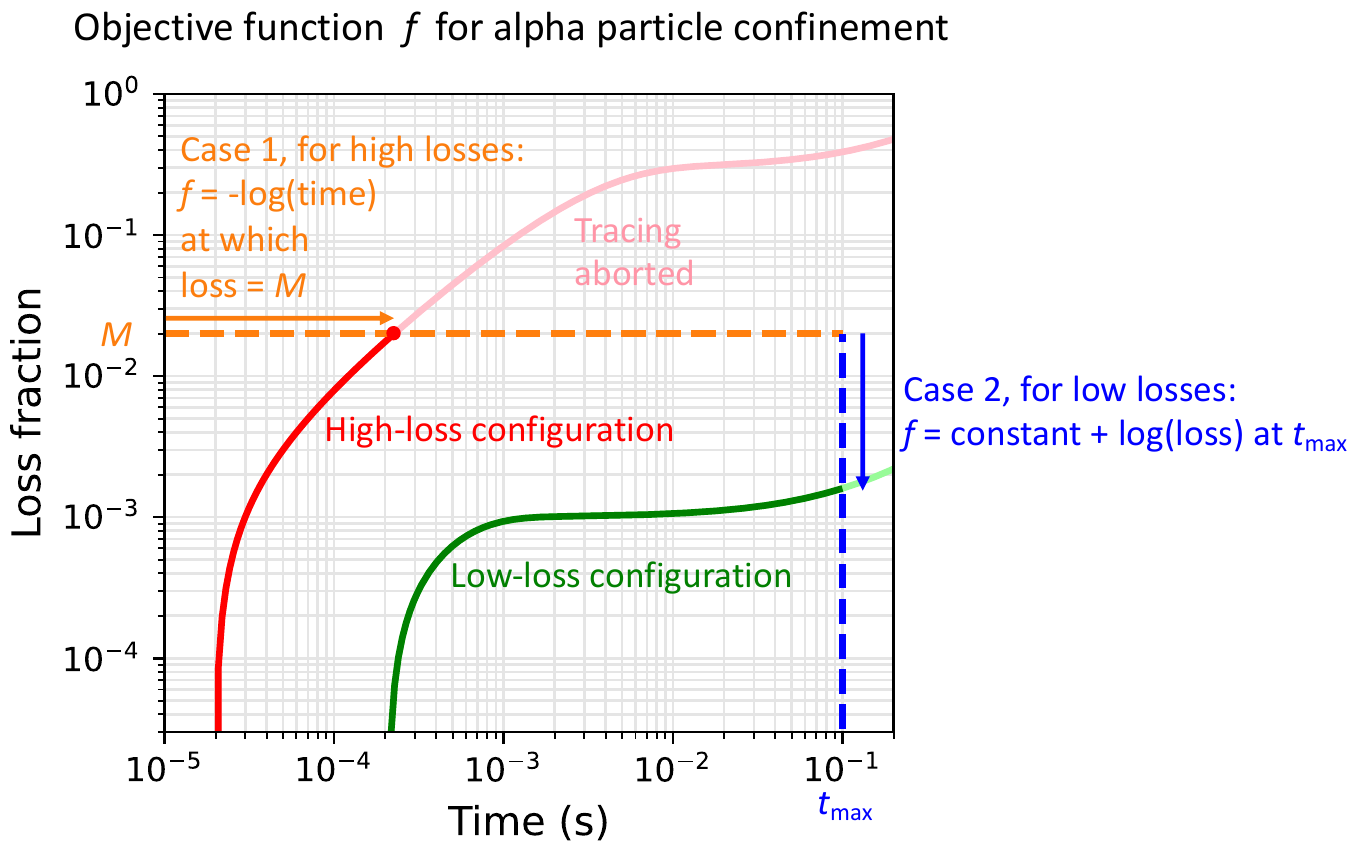}
 \caption{The objective function (\ref{eq:objective}) is chosen to allow early termination of the tracing for high-loss configurations, reserving computation time for the best-performing configurations.}
\label{fig:objective}
\end{figure}

An objective function is formulated so that the tracing can be terminated early for configurations with high losses, with particles traced only until the loss fraction exceeds a threshold $M$.
This approach effectively constitutes a multi-fidelity method, with more computation reserved for the best-performing configurations.
The objective function is illustrated in figure \ref{fig:objective}.
Let $L\left(t\right)$ denote the monotonically increasing energy loss as a function of time $t$.
For collisionless calculations, we consider a particle lost at time $t$ to constitute a loss $\exp(-t/\tau)$ of its initial energy where the constant $\tau$ is a slowing down time, as in \cite{bindel2023direct}.
Tracing is stopped as soon as $L\left(t\right)\ge M$, or at a prescribed maximum time $t_{max}$, whichever comes first.
The objective function $f$ is defined piecewise.
If $L$ ever exceeds $M$, we choose $f=-\log_{10}t_{M}$ where $t_{M}$ is the value of $t$
at which $L\left(t_{M}\right)=M$.
If $L\left(t_{max}\right)<M$, we want the objective to continue
to decrease if $L\left(t_{max}\right)$ decreases.
Therefore, if $L\left(t_{max}\right)<M$, we choose the objective to be $f=\log_{10}\left(L\left(t_{max}\right)+\epsilon\right)+C$
where $\epsilon$ is a small positive constant to prevent divergence
if $L=0$, and $C$ is a constant chosen to make $f$ continuous at
$L\left(t_{max}\right)=M$. To determine $C$, we equate the two cases
for $f$ at $L=M$, with the result
\begin{equation}
C=-\log_{10}\left(t_{max}\right)-\log_{10}\left(M+\epsilon\right).
\end{equation}
Hence the objective function is
\begin{equation}
f=\left\{ \begin{array}{cc}
-\log_{10}\left(t_{M}\right) & \mbox{ if }L\left(t_{max}\right)>M,\\
\log_{10}\left(L\left(t_{max}\right)+\epsilon\right)-\log_{10}\left(t_{max}\right)-\log_{10}\left(M+\epsilon\right) & \mbox{ if }L\left(t_{max}\right)\le M.
\end{array}\right.
\label{eq:objective}
\end{equation}
For the calculations that follow we choose $\tau=t_{max} = 0.1$ seconds (approximately a slowing-down time, \changed{and comparable to prior publications \cite{bader2019stellarator,paul2022energetic,goodman2024quasi}}), $M=2 \%$, and $\epsilon=1/25,000$.

If the equilibrium calculation fails to converge, the objective is set to 5.5.
This value is slightly higher than the objective for any equilibrium that does converge, in which case 2\% losses would not occur before $\sim 10^{-5}$ seconds, corresponding to $f=5$.
\changed{Resolution and tolerance parameters for the equilibrium and tracing were set to ensure that the alpha loss calculations were insensitive to refinement.}

There is no need to include a penalty or constraint that bounds the rotational transform $\iota$ away from 0, as has often been done in previous optimization formulations \cite{landreman2022magnetic,schuett2024exploring}.
If $|\iota|$ is too small, the equilibrium will not converge at the specified pressure, or the width of guiding-center trajectories will become large, leading to high alpha losses.
Thus the problem formulation (\ref{eq:objective}) automatically accounts for the requirement of sufficient $|\iota|$ for confinement.


\subsection{Optimization method}

To minimize the alpha-particle losses we employ Bayesian optimization, a derivative-free method that builds a probabilistic surrogate model (a Gaussian process) of the objective function from previous evaluations, representing not only the expected value of the objective but also the uncertainty in it.
An acquisition function is then optimized to select new evaluation points, balancing exploration of uncertain regions with exploitation of regions predicted to have favorable objective values.
In our optimizations we have tried both standard Bayesian optimization via the Ax library \cite{olson2025ax}, which uses the modification proposed in \cite{hvarfner2024vanilla} to scale to a higher number of dimensions, and the TURBO method for local Bayesian optimization \cite{eriksson2019scalable}.

\begin{figure}
 \centering
        \includegraphics[width=\textwidth]{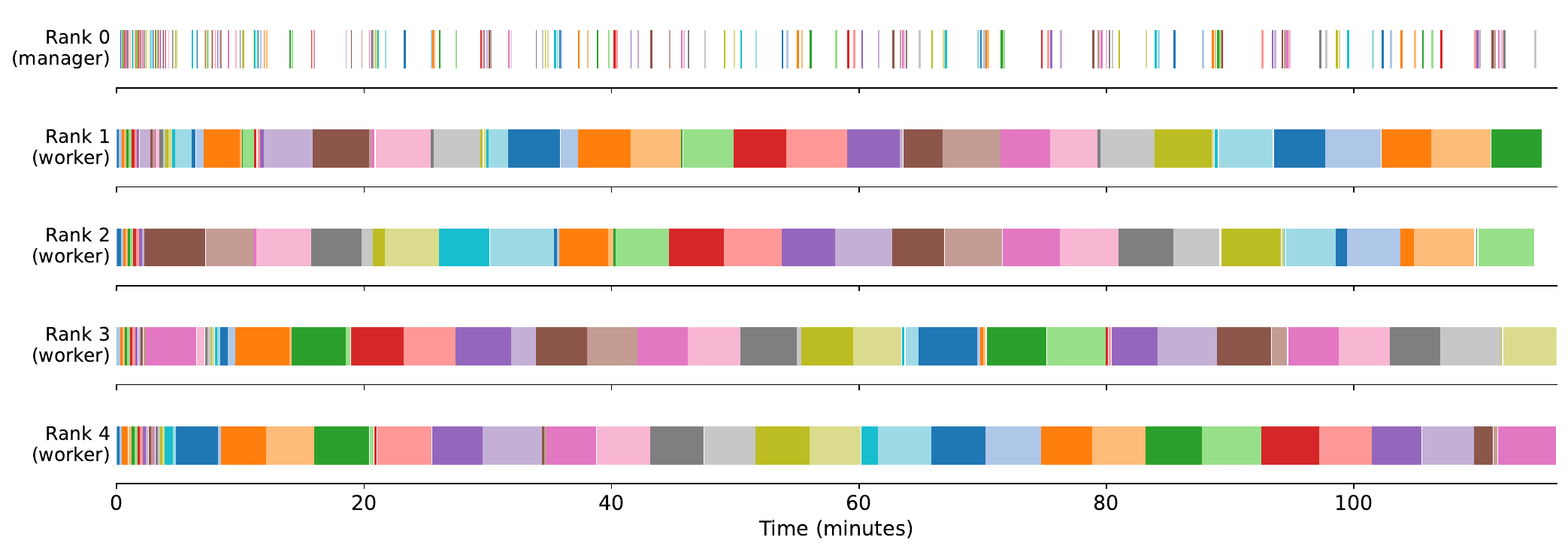}
 \caption{
Illustration of the asynchronous algorithm, showing the optimization of configuration 2 below.
For each of the four workers, corresponding to one of the GPUs on the compute node, each colored block indicates the evaluation of the objective function for one stellarator shape.
All workers are kept busy, and for shapes for which the equilibrium fails or with poor confinement, little time is used.
For the manager, which uses only CPU threads, the colored blocks show the time to generate samples from the Gaussian process.
 }
\label{fig:async}
\end{figure}

For both Ax and TURBO, we use an asynchronous method, meaning the objective function can be evaluated at multiple points in the parameter space concurrently by different workers, and each worker need not wait for others to finish before evaluating a new point.
\changed{The choice of an asynchronous method is linked to the objective function (\ref{eq:objective}), since this objective requires varied amounts of wallclock time to evaluate on different configurations, so a synchronous batched approach would result in some workers idling.}
We run on the Perlmutter computer, which has four GPUs per compute node, so we use four workers plus a manager process.
Each worker corresponds to one of the GPUs, and also gets $\sim 1/5$ of the CPU threads on the node.
No additional development is required to include multiple nodes for more GPUs if desired.
In addition there is a manager process using the remaining CPU threads on the node.
The manager generates new points to evaluate and sends them to the workers, and collects the results.
The asynchronous method is illustrated in figure \ref{fig:async}, which shows the optimization of configuration 2 described in the next section.
For each worker, each colored rectangle shows the evaluation of the objective for a different point in parameter space.
Some rectangles are thin, corresponding to points where the equilibrium calculation fails, or $2\%$ losses are reached early in the guiding center tracing, so the tracing is stopped.
The wide rectangles correspond to good points in parameter space, where the particles are traced all the way to $0.1$ seconds of physical time.
It can be seen that all workers are kept busy, with negligible time idling.
As soon as an objective function is returned to the manager, this new value is incorporated into the Gaussian process, and a new candidate point is generated.

\begin{figure}
 \centering
        \includegraphics[width=4in]{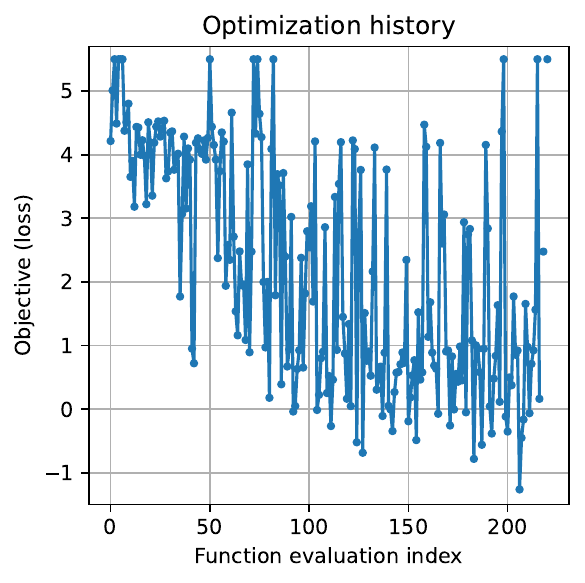}
 \caption{
Optimization history for configuration 2 (same as figure \ref{fig:async}).
 }
\label{fig:opt_history}
\end{figure}

The history of objective for the same optimization is shown in figure \ref{fig:opt_history}.
It can be seen that there are a few failures of the equilibrium calculation, corresponding to the points at $f=5.5$, but these points are rare.
The optimization makes steady progress, while balancing exploration and exploitation.

\subsection{Results}
\label{sec:results}

\begin{figure}
 \centering
 \setlength{\tabcolsep}{0pt}
 \begin{tabular*}{\textwidth}{@{\extracolsep{\fill}}ccccc@{}}
 \small Configuration 1 & \small Configuration 2 & \small Configuration 3 & \small Configuration 4 & \small Configuration 5 \\
 \includegraphics[width=0.196\textwidth]{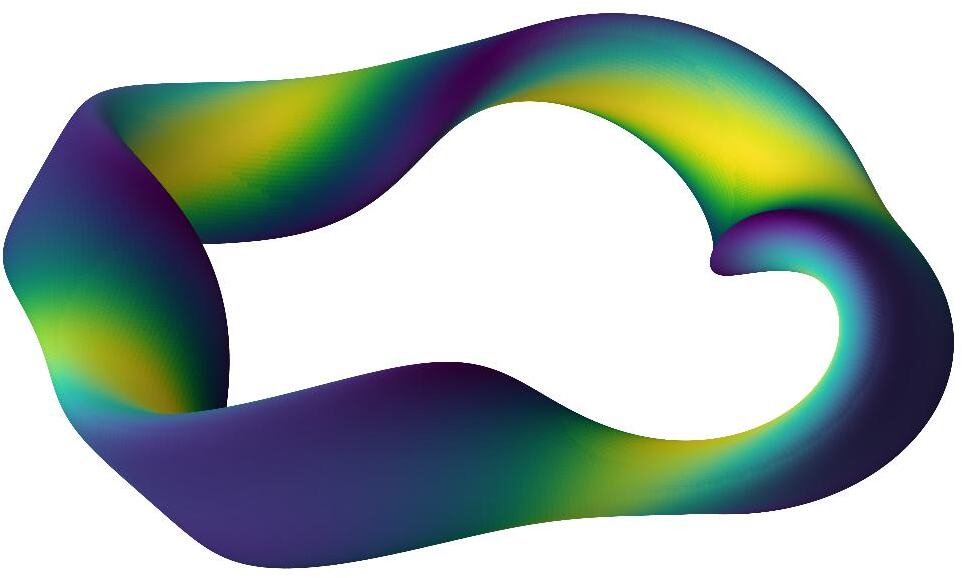} &
 \includegraphics[width=0.196\textwidth]{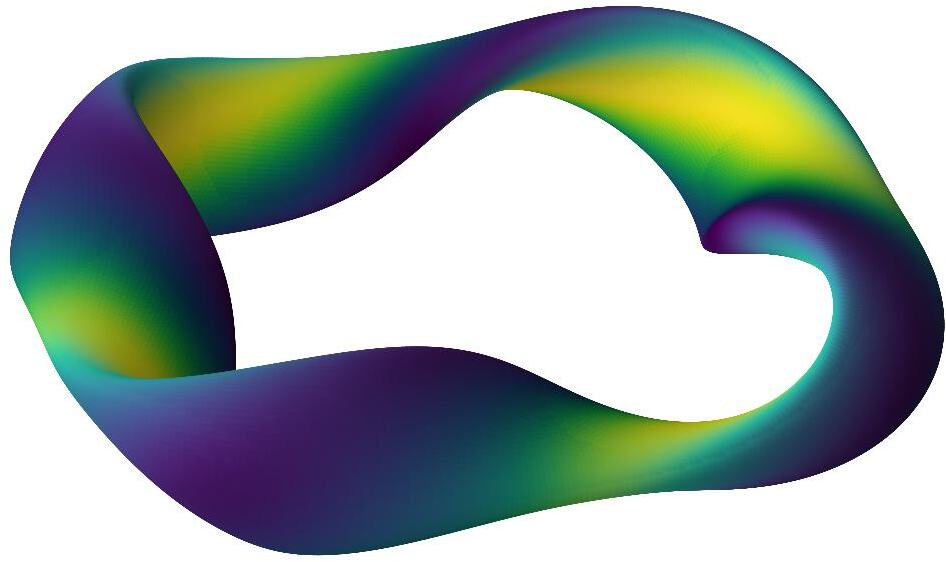} &
 \includegraphics[width=0.196\textwidth]{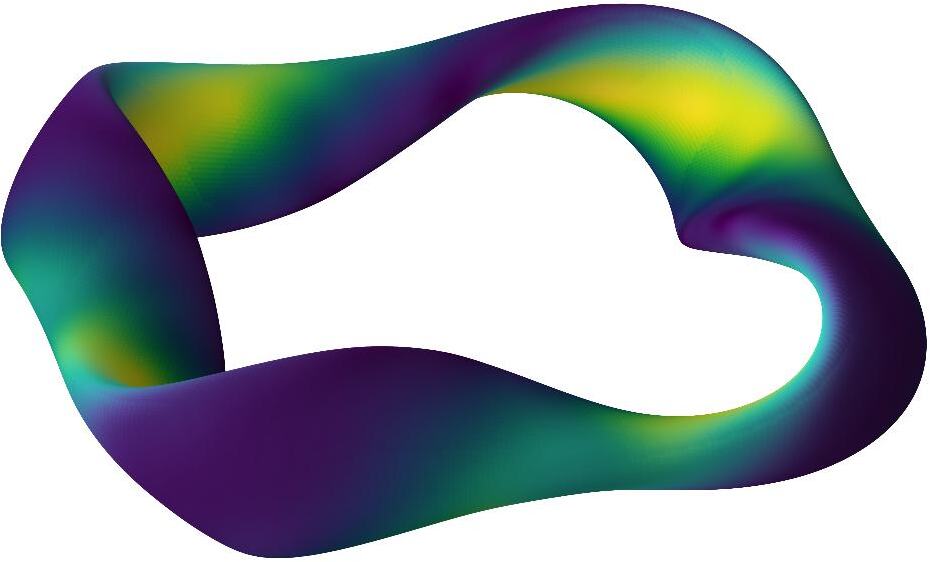} &
 \includegraphics[width=0.196\textwidth]{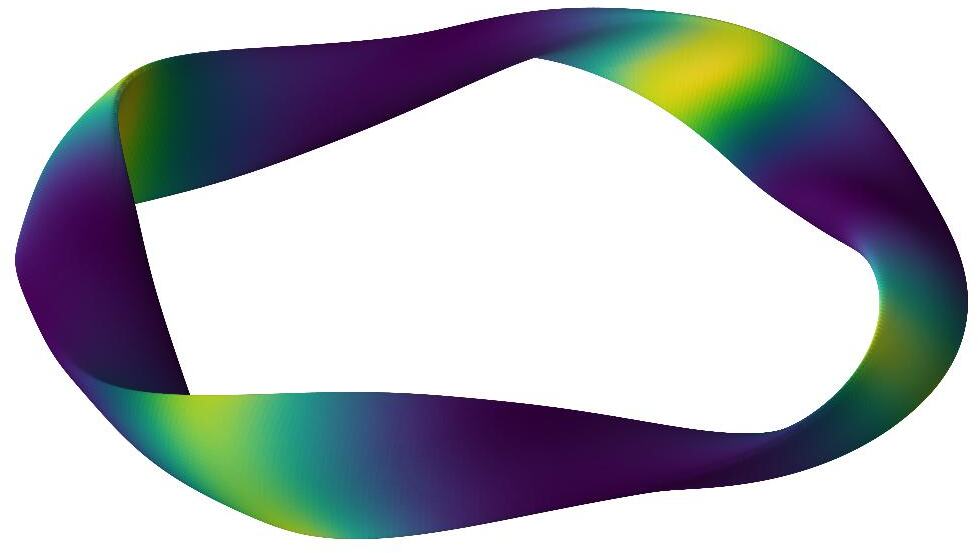} &
 \includegraphics[width=0.196\textwidth]{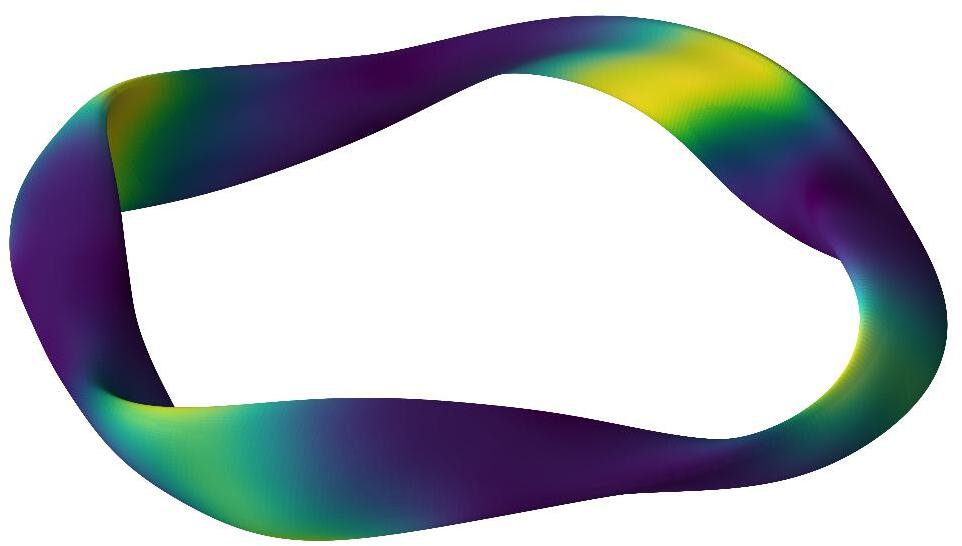} \\[-2pt]
 \raisebox{-0.5\height}[0pt][0pt]{\raisebox{4pt}{\includegraphics[width=0.196\textwidth]{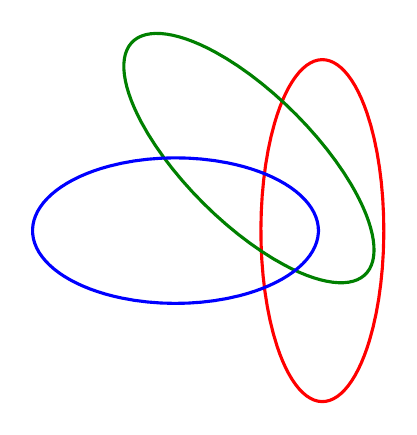}}} &
 \raisebox{-0.5\height}{\includegraphics[width=0.196\textwidth]{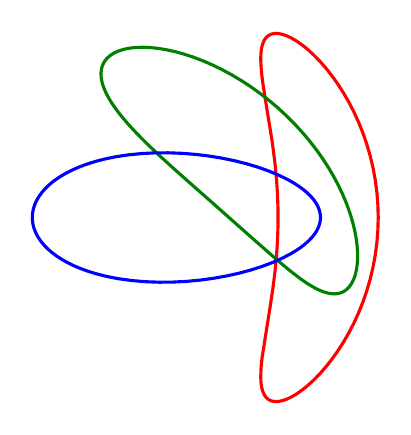}} &
 \raisebox{-0.5\height}{\includegraphics[width=0.196\textwidth]{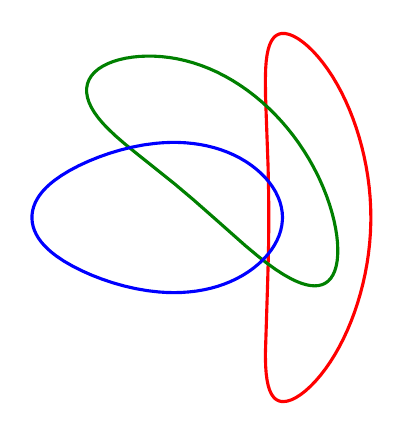}} &
 \raisebox{-0.5\height}{\includegraphics[width=0.196\textwidth]{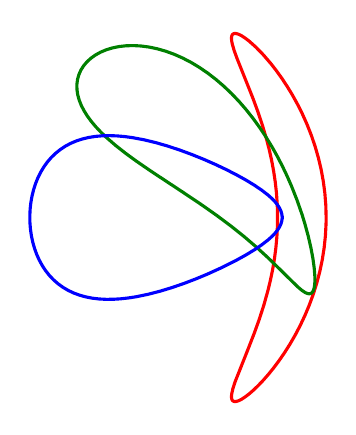}} &
 \raisebox{-0.5\height}{\includegraphics[width=0.196\textwidth]{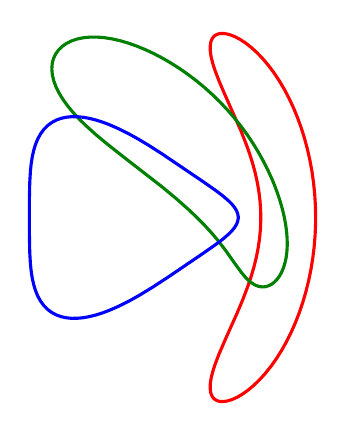}}
 \end{tabular*}
 \caption{The five new configurations obtained using Bayesian optimization.
 In the bottom row, cross-sections are shown at $\phi=0$ (red), $\pi/8$ (green), and $\pi/4$ (blue).
 }
\label{fig:configs}
\end{figure}

\begin{figure}
 \centering
        \includegraphics[width=4in]{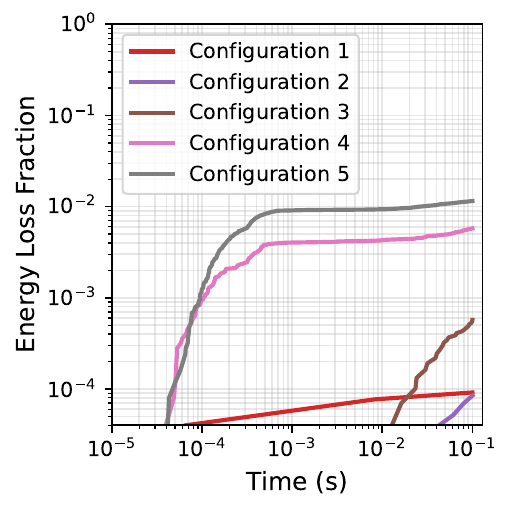}
 \caption{
Alpha-particle energy losses for the five optimized configurations, with particles initialized throughout the volume at the local fusion rate.
Losses are quite low, especially when Mercier stability is not enforced (configurations 1-3, loss $< 0.06\%$).
Even when Mercier stability is enforced (configurations 4-5) losses are $\le 1\%$.
 }
\label{fig:alpha_loss}
\end{figure}

Here we present the results of five optimizations, with the optimum from each referred to as configuration 1--configuration 5.
The MHD equilibria, along with scripts used for their optimization and the data to define the parameter spaces, are openly available on Zenodo \cite{zenodo}.
For all equilibria, we choose $n_{fp}=4$.
The boundaries for each geometry are shown in figure \ref{fig:configs}.
The energy loss versus time for each stellarator is displayed in figure \ref{fig:alpha_loss}.
The first three optimizations are unconstrained minimizations of the objective $f$ in eq (\ref{eq:objective}), fixing the aspect ratio at 6.
For configuration 1, we optimize in the Fourier space, with modes $|n| \le 1$ and Garabedian $m \in [0,2]$ (corresponding to standard Fourier modes $|m| \le 1$).
This space has 7 degrees of freedom, and the optimization algorithm was TURBO.
Remarkably, even though the boundary shaping is minimal (a rotating ellipse with axis torsion), this geometry still achieves an alpha energy loss $< 0.01 \%$.
Configuration 2 was optimized using Ax in the Fourier parameter space with more degrees of freedom: $|n| \le 2$ and Garabedian $m \in [-1, 3]$ (corresponding to standard $|m| \le 2$), for a total of 23 degrees of freedom.
Configuration 3 was optimized with Ax in the PCA parameter space, varying the first 20 principal components.
For the remaining two configurations, a constraint for Mercier stability was included in Ax.
Details of the normalization for this constraint are given in \ref{sec:mercier}.
It was much harder to achieve good confinement when this constraint was included, so the aspect ratio was increased to 10.
Configuration 4 was generated using the same Fourier parameter space as configuration 2: $|n| \le 2$ and Garabedian $m \in [-1, 3]$ for 23 degrees of freedom.
Configuration 5 was optimized using the PCA space, varying 25 principal components.
The volume-averaged $\beta$ values for the five configurations are $1.5\%$, $1.4\%$, $1.5\%$, $2.9\%$, and $1.9\%$ respectively, reflecting variation in the average $B$ at fixed $B_{\max}$ and pressure.

\begin{figure}
 \centering
        \includegraphics[width=3in]{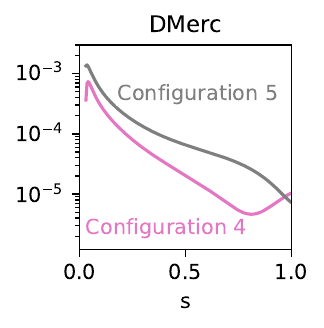}
 \caption{
Demonstration of Mercier stability ($D_{Merc} > 0$) for configurations 4 and 5.
 }
\label{fig:mercier}
\end{figure}

The alpha-particle energy losses shown in figure \ref{fig:alpha_loss} are quite low for all five configurations.
This is especially true when Mercier stability is not enforced (configurations 1-3), and the losses are $< 0.06\%$.
Even when Mercier stability is enforced (configurations 4-5) losses are $\le 1\%$, in line with e.g. \cite{hegna2025infinity}.
Figure \ref{fig:mercier} shows that configurations 4-5 are indeed Mercier-stable.

\changed{
For configurations 1-2, each parameter is in the interior of the space: $\min(u) > 0.30$, $\max(u) < 0.72$.
However configurations 3-5 each have one parameter close to the boundary of the space: $\max(x)=$0.96, 0.97, 0.96 respectively.
This finding suggests that there may be nearby optima which are not expressible in the spaces used for these three optimizations, leaving room for improvement in the choice of parameter spaces.
}

\begin{figure}
 \centering
        \includegraphics[width=4in]{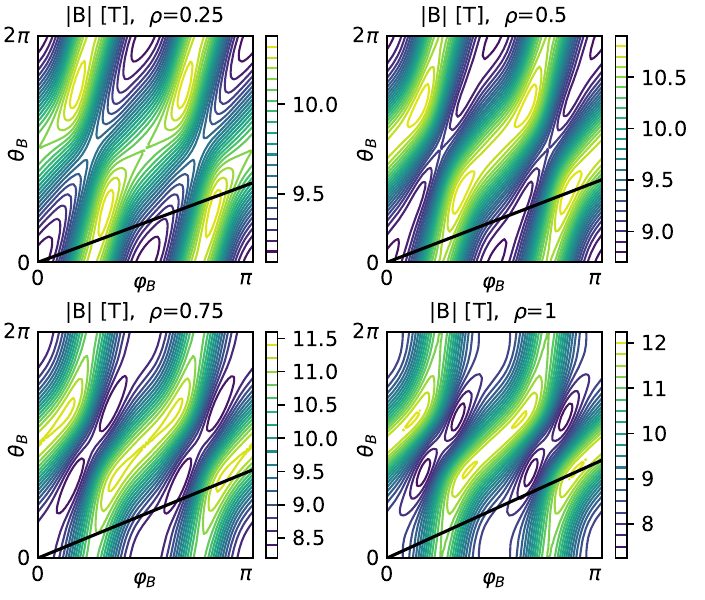}
 \caption{
Magnetic field strength on four flux surfaces of configuration 1, as functions of the 
 poloidal and toroidal Boozer angles.
 Two field periods are shown, and the black diagonal line indicates the field direction.
 }
\label{fig:B_contours_1}
\end{figure}

\begin{figure}
 \centering
        \includegraphics[width=4in]{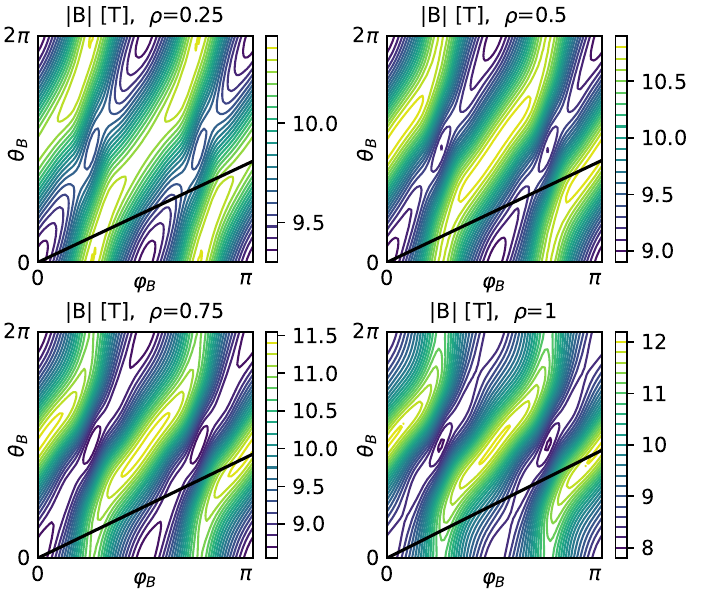}
 \caption{
Magnetic field strength on four flux surfaces of configuration 2, as functions of the 
 poloidal and toroidal Boozer angles.
 Two field periods are shown, and the black diagonal line indicates the field direction.
 }
\label{fig:B_contours_2}
\end{figure}

\begin{figure}
 \centering
        \includegraphics[width=4in]{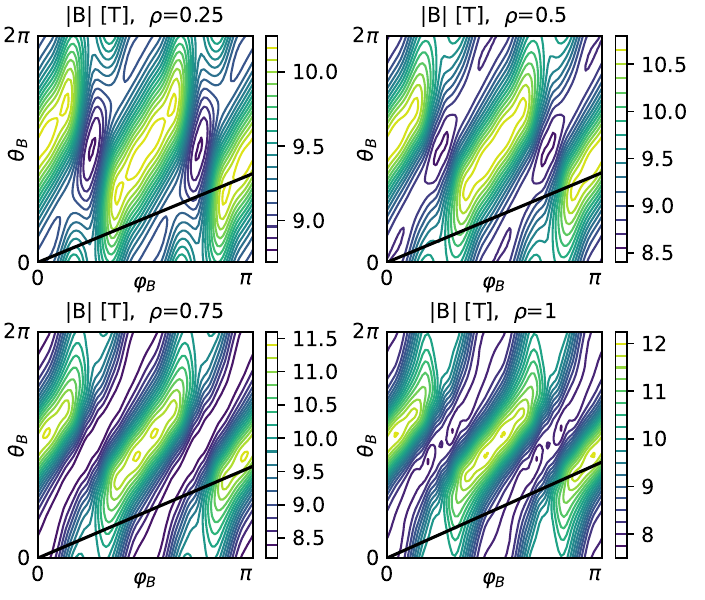}
 \caption{
Magnetic field strength on four flux surfaces of configuration 3, as functions of the 
 poloidal and toroidal Boozer angles.
 Two field periods are shown, and the black diagonal line indicates the field direction.
 }
\label{fig:B_contours_3}
\end{figure}

\begin{figure}
 \centering
        \includegraphics[width=4in]{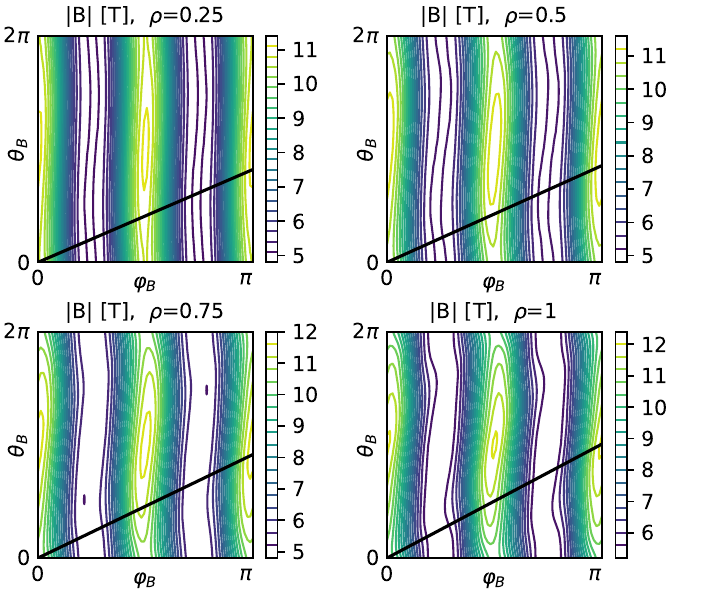}
 \caption{
Magnetic field strength on four flux surfaces of configuration 4, as functions of the 
 poloidal and toroidal Boozer angles.
 Two field periods are shown, and the black diagonal line indicates the field direction.
 }
\label{fig:B_contours_4}
\end{figure}

\begin{figure}
 \centering
        \includegraphics[width=4in]{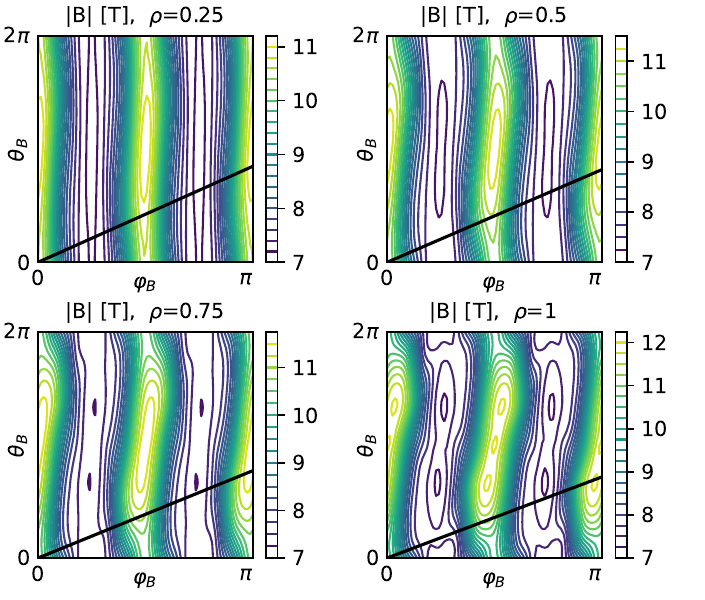}
 \caption{
Magnetic field strength on four flux surfaces of configuration 5, as functions of the 
 poloidal and toroidal Boozer angles.
 Two field periods are shown, and the black diagonal line indicates the field direction.
 }
\label{fig:B_contours_5}
\end{figure}

The field strength $B$ as a function of Boozer angles is shown for these five optimized configurations in figures \ref{fig:B_contours_1}-\ref{fig:B_contours_5}.
In each case, results are shown for four flux surfaces, and a black line on each panel indicates the field direction.
Configurations 1-2 are somewhat reminiscent of quasi-helical symmetry, or omnnigenity with helically closed contours, at least for intermediate-$B$ contour levels, but deviations from symmetry are large.
In configuration 3, the $B$ contours for $B> 10$ T resemble parallelograms, as appear in piecewise-omnigenous fields \cite{velasco2025exploration}.
Meanwhile the contours for $B < 10$ T are closed helically, so overall the contours resemble the hybrid omnigenous-piecewise-omnigenous fields in Refs \cite{liu2026optimization, velasco2026combination}.
Configurations 4-5 look similar to recent quasi-isodynamic designs \cite{sanchez2023quasi, goodman2024quasi, lion2025stellaris, hegna2025infinity}.
In all five of the new configurations, the contours of $B$ near $B_{\max}$ do not link the flux surface poloidally, toroidally, or helically, violating omnigenity.
Only in configuration 4 do the contours of $B$ near $B_{\min}$ link the surface at each radius.
In the other four configurations, the contours of $B$ near $B_{\min}$ close without linking the surface at least for some radii, violating omnigenity.
Overall, then, none of the configurations are close to being quasisymmetric, and while configurations 4-5 may be somewhat close to quasi-isodynamic, they depart from omnigenity significantly near $B_{\max}$.
It is remarkable that such good confinement of alpha particles is possible when the deviations from QS and QI are so large.

It is interesting to compare the new configurations 4-5 to the configurations from ConStellaration in the input dataset.
These ConStellaration configurations were optimized for QI and magnetic well, so one might expect them to be similar to our geometries 4-5.
The ConStellaration configurations in our dataset were each scaled to the same minor radius and $B_{\max}$ as described in section \ref{sec:profiles}, \changed{without altering the aspect ratios}, and guiding center trajectories were initialized using the fusion reaction rate with the same density and temperature profiles as in section \ref{sec:profiles}, \changed{following particles for $t=0.1$ seconds.}
\changed{These calculations were done both taking the pressure in the MHD equilibria to be zero (as was done in the ConStellaration optimizations) and using our pressure profile (\ref{eq:pressure_profile}), to see which gave lower losses.
Similarly, the loss calculations were done both at the original values of $n_{fp}$ and with $n_{fp}$ changed to 4, to see which gave lower losses.}
All these ConStellaration configurations had energy losses of $> 15\%$; none came close to the $\le 1\%$ losses of the new geometries from the Bayesian approach.

\changed{
It is possible that the optimization results and the confinement in these configurations may be modified by addition of self-consistent bootstrap current.
While a full calculation of this effect is left for future work, Table 3 of \cite{schmitt2025magnetohydrodynamic} shows the bootstrap current only alters the edge $\iota$ by 0.001 for similar plasma parameters in a configuration similar to configurations 4-5 here.
}

\begin{figure}
 \centering
        \includegraphics[width=3.3in]{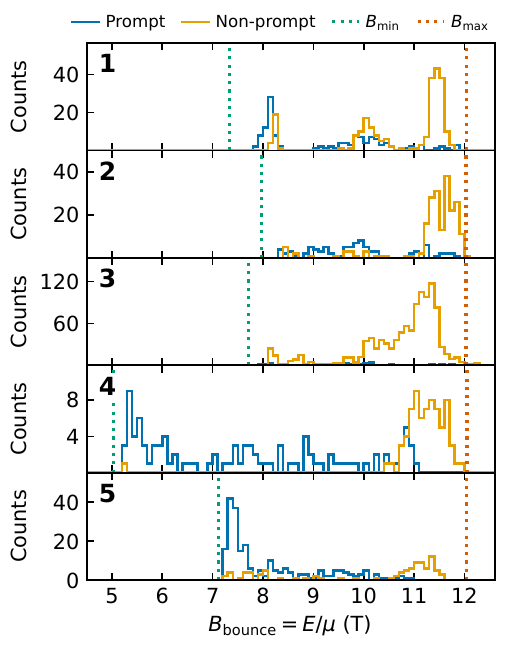}
 \caption{
Dependence of the losses on pitch-angle.
The five panels show data for configurations 1-5.
Prompt losses are those for $t<10^{-3}$ sec.
 }
\label{fig:pitch_angles}
\end{figure}

\changed{
The distribution of the conserved pitch-angle parameter $E/\mu$ is shown in figure \ref{fig:pitch_angles} for the particles lost from the five configurations.
Here, $E=v^2/2$ is the kinetic energy and $\mu=v_{\perp}^2/(2B)$ is the magnetic moment.
For configurations 1-3, where the loss rates are lowest, a total of $10^6$ guiding center trajectories were traced, to collect better statistics.
In all configurations, losses of passing particles ($E/\mu > B_{\max}$) are insignificant.
Losses are categorized as prompt ($t < 10^{-3}$) or non-prompt ($t \ge 10^{-3}$).
In all five configurations, non-prompt losses occur predominantly for barely trapped particles ($E/\mu \sim B_{\max}$).
This trend may be consistent with slow transport from particles that transition between trapping classes \cite{beidler2001stochastic}.
For configurations with prompt losses, they tend to be more pronounced for deeply trapped pitch angles ($E/\mu \sim B_{\min}$), except in configuration 2. 
Further analysis would be valuable to understand the loss mechanisms in greater detail; we leave this to future work.
}

\begin{figure}
 \centering
        \includegraphics[width=3in]{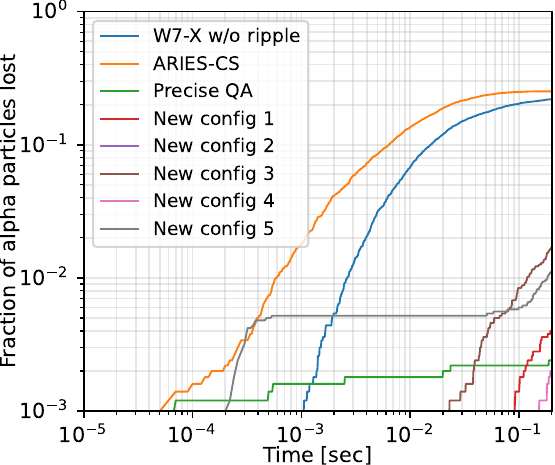}
 \caption{
Collisionless losses of alpha particles for the new configurations and several well-known ones using the alternative scaling and initialization from \cite{landreman2022magnetic}.
Here, all configurations are scaled to the minor radius and volume-averaged $B$ of ARIES-CS, and particles are initialized at $s=0.25$.
 }
\label{fig:simple}
\end{figure}

For additional context, in figure \ref{fig:simple} we show alpha-particle losses for the new configurations using the alternative scaling and initialization from \cite{landreman2022magnetic}.
Here, all configurations are scaled to the minor radius and volume-averaged $B$ of ARIES-CS, $a=1.7$ m and $\langle B \rangle  = 5.9$ T, and particles are all initialized at the surface with normalized toroidal flux $s=0.25$.
Guiding center trajectories of 5,000 particles are followed without collisions for 0.2 seconds.
Several well known configurations are also shown in the figure, including a stage-1 variant of W7-X which has no coil ripple, for a fair comparison.
It can be seen that the new configurations all have losses $< 2\%$ with this normalization, over an order of magnitude better than W7-X and ARIES-CS.

\begin{figure}
 \centering
        \includegraphics[width=3in]{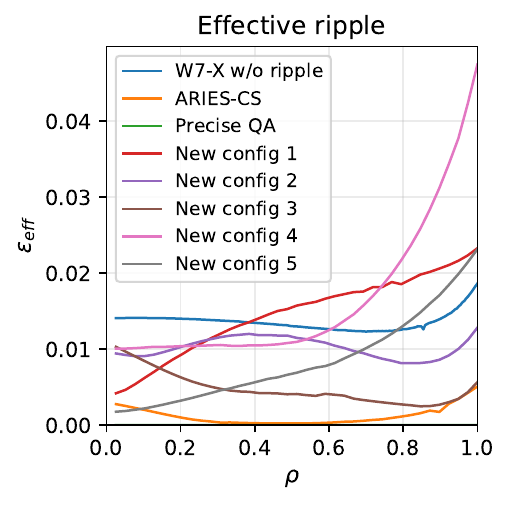}
 \caption{
Profiles of effective ripple for the new configurations here as well as several well-known stellarators.
Even without optimization of $\epsilon_{eff}$ or optimization for QS or omnigenity, the new configurations have $\epsilon_{eff} < 1\%$ in the core, better than W7-X, which is likely good enough.
 }
\label{fig:eps_eff}
\end{figure}

One pattern seen in \cite{bindel2023direct} and again observed here is that optimization of fast-ion losses automatically results in good neoclassical \emph{thermal} transport.
This pattern can be seen in figure \ref{fig:eps_eff}, which displays the effective ripple $\epsilon_{eff}$ for the new configurations, as well as for several well-known stellarators.
The new configurations optimized for fast-ion transport all have $\epsilon_{eff} < 1\%$ in the core, better than W7-X, which is likely good enough \cite{alonso2022physics}.
Away from the core, $\epsilon_{eff}$ is less important due to the very peaked $\propto T^{9/2}$ behavior of the $1/\nu$-regime neoclassical heat flux.
It is not surprising that optimization of fast-ion losses leads to reduced $\epsilon_{eff}$, since both are associated with radial magnetic drift.
While fast-ion losses also depend on other factors such as poloidal drifts, the correlation with $\epsilon_{eff}$, while modest, is good enough that no additional objective or constraint for $\epsilon_{eff}$ is required in practice.
Good fast-ion confinement appears to be much more restrictive a condition than $\epsilon_{eff} < 1\%$, for many configurations such as ARIES-CS are known to have sufficiently small $\epsilon_{eff}$ yet high alpha-particle losses.

\begin{figure}
 \centering
        \includegraphics[width=3in]{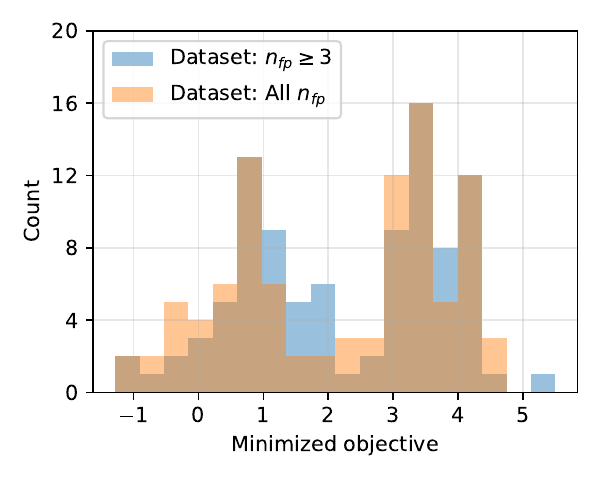}
 \caption{
Distribution of minimum values of the objective function for 192 independent optimizations, showing the dependence on which data were used to define the space.
The results are insensitive to whether or not $n_{fp}=1$ and 2 configurations were included.
 }
\label{fig:histogram_vs_nfp}
\end{figure}

To explore the sensitivity of the optimizations to which configurations are included in the data, 192 unconstrained optimizations were run, half using the full dataset and half including only the configurations with $n_{fp} \geq 3$.
For each choice, other parameters such as the aspect ratio and dimensionality of the parameter space were varied equally across the optimizations.
The distributions of final objective values are shown in figure \ref{fig:histogram_vs_nfp}.
There is not a significant difference in the distributions betwen the $n_{fp} \geq 3$ and all-$n_{fp}$ datasets, with both choices achieving similarly small values of the objective, $\sim -1$.
This finding suggests that optimizations may not be overly sensitive to the choice of data, at least in some cases.

Another test of sensitivity to the dataset was performed by running additional Mercier-stability-constrained optimizations (as for configurations 4 and 5) with a reduced dataset.
In this experiment, all QI-like configurations, including all ConStellaration configurations and W7-X variants, were removed from the data.
This was done to see if it was still possible to arrive at QI-like optima with no QI configurations in the input data.
None of these optimizations produced results with as good confinement as configurations 4 and 5, and while one QI-like result was found, most optima had $B$ contours that primarily closed helically instead of poloidally. 
From this result we conclude that the selection of data can affect the optima obtained. 
Therefore users of these methods should try different datasets and weights to see which result in the best optimizations.
Note that although a dataset and weights must be chosen to define the parameter space, the resulting optimized configurations are ultimately represented by standard Fourier amplitudes that are independent of that choice.

\section{Discussion and conclusions}
\label{sec:conclusions}

In this paper we have demonstrated two new effective parameter spaces for stellarator optimization.
Both methods use a set of known boundary shapes, and exploit quantile transformation so all degrees of freedom have similar scale and self-evident bound constraints.
The two spaces are simultaneously capable of representing strongly shaped boundaries while minimizing self-intersections so a large fraction of the space is usable.
Both spaces significantly push the Pareto front for the trade-off between these competing concerns compared to a traditional Fourier description, as summarized in figure \ref{fig:Pareto}.
The PCA space allows the additional benefit of dimensionality reduction.
Due to all these features, these spaces are promising for global and/or derivative-free optimization algorithms.
As a demonstration, we used both spaces for Bayesian global optimization with guiding-center tracing of alpha particle losses inside the objective.
The results add to the evidence that acceptably low \changed{collisionless} alpha losses can be achieved in fields that are far from being quasisymmetric or quasi-isodynamic, \changed{at least for the device sizes and field strengths considered here}.

There are many directions for future research, both related to the parameter spaces and to the confinement optimizations.
For the former, rather than applying independent quantile transformations to each parameter, other maps from the data distribution to the unit hypercube could be devised to account for correlations between the parameters.
\changed{Ideally, a uniform distribution on the parameter space would represent a prior matching the data distribution including its nonlinear correlations, not only matching certain marginal distributions as done here.}
Other dimensionality reduction methods could be explored, which may enable compression of the space of useful stellarator boundaries to a lower number of dimensions.
More systematic studies could be conducted of how the spaces depend on the choice of data.
For instance, it is not yet clear whether or not it is preferable to pool data from multiple values of $n_{fp}$ and aspect ratio, trading off more data vs more targeted data.
The parameter spaces here could also be applied in derivative-based optimization, as the quantile and principal component transformations are differentiable, and in single-stage optimization.

Many extensions are also possible for the specific application of alpha-particle confinement optimization.
Importance sampling and other techniques could be used to improve the tracing statistics \cite{law2022accelerating}.
Collisions could be included in the tracing.
Optima could be refined using local optimization methods.
\changed{
Sensitivity analysis should be carried out to determine the extent to which the minima in the objective function are broad or narrow.
Stochastic optimization, based on the expected value of the Gaussian process surrogate over a region surrounding each evaluation point, could be used to steer the optimizations towards optima that are not too narrow (i.e. not too sensitive) \cite{glas2022global}.
}
More physics objectives can be included in the optimization, such as ballooning stability, bootstrap current, and \changed{the magnetic gradient scale length} $L_{\nabla B}$.
A particularly interesting extension would be to directly include turbulent heat fluxes from nonlinear gyrokinetic simulations.
The optimization methods in this paper should be well suited to the noisy and non-differentiable fluxes from such simulations, and with adiabatic electrons the cost of these calculations is now within a factor of $\sim 3$ of the guiding-center tracing used here \cite{landreman2025does}.


\ack{
We gratefully acknowledge conversations about this work with 
David Bindel,
Michael Churchill,
Neil Dhir,
David Erikkson,
Hengqian Liu,
Misha Padidar,
Sihwa Park,
Nathaniel Smith,
Jos\'e Luis Velasco,
and
Caoxiang Zhu.
	}

\funding{
This work was supported by the US Department of Energy under contract DE-AC02-09CH11466 (High-fidelity Digital Models for Fusion Pilot Plant Design, StellFoundry) and contract DE-FG02-93ER54197. 
R.C. was supported by Greg Hammett’s DOE Distinguished Scientist Fellow award. 
This research used resources of the National Energy Research Scientific Computing Center (NERSC), a Department of Energy Office of Science User Facility using NERSC awards FES-mp217 and FES-m4505 for 2025-2026.	
}


\data{
Data and software associated with this paper are openly available at \url{https://doi.org/10.5281/zenodo.20733436} \cite{zenodo}.
}


\appendix

\section{Details of the data preparation}
\label{sec:data_prep_details}

Here we give more details of the data selection and preparation.
As described in section \ref{sec:data_prep}, the boundary shapes are drawn from Kappel et al \cite{kappel2024magnetic}, QUASR \cite{giuliani2024direct, giuliani2024comprehensive}, and ConStellaration \cite{cadena2025constellaration}.
Of the configurations listed in \cite{kappel2024magnetic}, we exclude ITER since it is not a stellarator.
We also add configurations A and B from \cite{bindel2023direct}.

We select a subset of the plasma shapes from QUASR to include as follows.
For each $n_{fp}\in [2,3,4,5]$, we loop over aspect ratio bins $\{[2,3], [3,4], \ldots, [9,10]\}$ and each unique value of $\iota$ used in the QUASR optimizations. In each of these bins, we choose the two configurations with lowest quasi-axisymmetry error and the two with lowest quasi-helical symmetry error.
This procedure gives 622 configurations.
We then apply spectral condensation to the boundaries, and scale each configuration in size and field strength as described in section \ref{sec:profiles}, using the pressure profile of equation (\ref{eq:pressure_profile}).
We keep only the boundary shapes for which \texttt{vmec++} then converges to a normalized force residual of $10^{-11}$.
This condition leaves 187 boundaries from QUASR to include in our final dataset.

For the ConStellaration data, we first apply basic filters to the set of $\langle \beta \rangle=2\%$ configurations on Hugging Face as follows:
$A \le 10$, elongation $\le 10$, $|\iota| \ge 0.4$ on axis and at the boundary, stable magnetic well, $B_{max} / B_{min} \le 2$, and $L_{\nabla B}/a \ge 0.58$ (where 0.58 is the worst value for any configuration in \cite{kappel2024magnetic}).
This leaves 1747 configurations.
We then apply spectral condensation to the boundaries, and scale them in size and field strength as described in section \ref{sec:profiles}, using the pressure profile of equation (\ref{eq:pressure_profile}).
Of the resulting equilibria, we select the ones that are Mercier-stable or nearly so, to increase the representation of stable configurations in our final dataset.
Only 14 of the filtered and scaled ConStellaration configurations are Mercier-stable for $s > 0.2$, so we weaken the selection criterion to allow moderate instability, $\min_{s>0.2} (D_{Merc} / N) > -0.01$  where $N$ is the normalization described in \ref{sec:mercier}.
This results in 38 accepted ConStellaration boundaries.


\section{Spectral condensation}
\label{sec:condensation}

To reduce the redundancy in the poloidal angle and systematize the poloidal angle across the data, spectral condensation is applied to each boundary shape in the input data.
This was especially important for the QUASR configurations, since the poloidal angle used in the QUASR optimizations was the Boozer angle, which does not typically give condensed Fourier series.
We observed that condensation reduced the change to $(R,Z)$ values when boundaries from the input dataset were projected onto a small number of principal components.

The condensation algorithm applied was different from the one described in \cite{hirshman1985optimized}.
Letting $\theta$ denote the original poloidal angle, a single-valued shift $\lambda(\theta,\phi)$ is applied to define a new poloidal angle $\vartheta=\theta+\lambda$.
Any particular choice of $\lambda(\theta,\phi)$ results in a new set of Fourier amplitudes $R'_{m,n}$ and $Z'_{m,n}$ defined with respect to the new angle $\vartheta$,
\begin{equation}
R(\vartheta,\phi) = \sum_{m,n}R'_{m,n}\cos(m\vartheta-n_{fp} n\phi),
\;\;\;
Z(\vartheta,\phi) = \sum_{m,n}Z'_{m,n}\sin(m\vartheta-n_{fp} n\phi),
\label{eq:Fourier_new}
\end{equation}
in contrast to the un-primed $R_{m,n}$ and $Z_{m,n}$ defined with respect to the old angle $\theta$ as in (\ref{eq:Fourier}).
The aim of condensation is to find the optimal $\lambda$ to minimize a spectral width objective
\begin{equation}
	f_w = \frac{1}{2} \sum_{m,n} \left( \frac{ (R'_{m,n})^2 +  (Z'_{m,n})^2 }{ a^2 } \right) (m^2 + n^2)^{p}.
\end{equation}
Here $a$ is the minor radius and the exponent $p$ is chosen to be 2.
Whereas powers of $n$ were not included in the spectral width objective in \cite{hirshman1985optimized}, we find that including $n$ is required to remove unnecessary toroidal variation in the origin of $\theta$.
(Consider e.g.~an axisymmetric circular-cross-section torus in which the origin of $\theta$ is shifted to depend on $\phi$.)
Inequality constraints are included in the optimization to ensure that the surface does not change in shape by more than a specified fraction $\epsilon$ of the minor radius.

In our implementation, values $R_j$ and $Z_j$ are first computed on the surface using a tensor-product grid of $N$ points in $(\theta,\phi)$ covering a half field period, with $j=1\ldots N$ a flattened index over both dimensions.
The $\phi$ grid is shifted by half of the $\phi$ grid spacing so the mean over the $\phi$ grid corresponds to $(2\pi)^{-1}\int_0^{2\pi} d\phi$.
The angle shift $\lambda$ is discretized using
\begin{equation}
\lambda(\theta,\phi) = \sum_{m,n}\lambda_{m,n}\sin(m\theta-n_{fp} n\phi),
\end{equation}
so the amplitudes $\lambda_{m,n}$ become the degrees of freedom for optimization.
For any specific values of these parameters, the new Fourier amplitudes are computed from
\begin{eqnarray}
Z'_{m,n} &=& \frac{1}{2\pi^2}\int_0^{2\pi} d\vartheta \int_0^{2\pi} d\phi\, Z\sin(m\vartheta-n_{fp} n \phi) \\
&=& \frac{1}{2\pi^2} \int_0^{2\pi} d\theta \int_0^{2\pi} d\phi\left( 1 + \frac{\partial\lambda}{\partial\theta}\right)Z\sin(m[\theta+\lambda]-n_{fp} n \phi) \nonumber \\
&\approx & \frac{2}{N}\sum_{j=1}^N \left( 1 + \frac{\partial\lambda}{\partial\theta}(\theta_j,\phi_j)\right)Z(\theta_j,\phi_j) \sin(m[\theta_j+\lambda(\theta_j,\phi_j)]-n_{fp} n \phi_j), \nonumber
\end{eqnarray}
with a similar expression for $R'_{m,n}$.
From these new Fourier amplitudes and the original $(\theta,\phi)$ grid values, new $(R,Z)$ values denoted $(R'_j, Z'_j)$ are then evaluated using the new Fourier series (\ref{eq:Fourier_new}).
The optimization problem to solve becomes $\min f_w$ subject to $|R_j - R'_j| \le \epsilon a$, $|Z_j - Z'_j| \le \epsilon a$ for all points $j$.
This constrained optimization problem is solved using the SLSQP implementation in \texttt{scipy}, with gradients obtained using automatic differentiation, and $\epsilon=10^{-3}$.
For maximum reliability, first the $|m| \le 1$ and $|n| \le 1$ modes of $\lambda_{m,n}$ are optimized, then the $|m| \le 2$ and $|n| \le 2$ modes, etc., and exponential spectral scaling is applied \cite{jang2026exponential}.
The values of $R'_{m,n}$ and $Z'_{m,n}$ at the optimal $\lambda_{m,n}$ become the new Fourier amplitudes to define the surface shape.


\section{Normalization of the Mercier stability metric}
\label{sec:mercier}


The condition for Mercier stability can be expressed as $D_{Merc}\left(\rho\right)>0$
for all $\rho$, where $D_{Merc}$ is given in eq (4.1) of \cite{landreman2020magnetic} and is computed in the equilibrium codes VMEC and DESC. However,
$|D_{Merc}|$ typically decreases strongly with minor radius (as can be seen in figure \ref{fig:mercier} and \cite{landreman2020magnetic}), such that
a constraint $\min_{\rho}D_{Merc}>0$ would weigh radii close to the
magnetic axis much more than radii at the edge. Moreover, $D_{Merc}$ scales
with pressure, and is not invariant if an MHD equilibrium
is scaled in size or field strength. Therefore, to make optimization
problems involving Mercier stability better scaled, here we derive
a radius-dependent normalization $N\left(\rho\right)$ such that $D_{Merc}/N$
will be (1) only weakly dependent on radius, (2) approximately
independent of the pressure, and (3) independent of the device size
and field strength when the MHD equilibrium is scaled. 

In practice, $D_{Merc}$ is usually dominated by the terms $D_{well}$
and $D_{Geod}$ (see eq (4.19)-(4.20) in \cite{landreman2020magnetic}), making
it reasonable to treat the $V''$ term in $D_{well}$ as a measure
of the size of $D_{Merc}$ overall. Here $V\left(\psi\right)$ is
the volume enclosed by a flux surface, $V''=d^{2}V/d\Psi^{2}$, and
$\Psi$ is the toroidal flux. The definition of $D_{well}$ is
eq (4.19) in \cite{landreman2020magnetic}:

\begin{equation}
D_{well}=\mu_{0}\frac{dp}{d\Psi}\left(\int dS\frac{B^{2}}{\left|\nabla\Psi\right|^{3}}\right)s_{\psi}\frac{d^{2}V}{d\Psi^{2}}+...
\end{equation}
where $s_{\psi}$ is the sign of $\Psi$, and $\int dS$ is an integral with respect to surface area. 
Our objectives will be acheived if we can find a radius-dependent
normalization $N\left(\rho\right)$ such that
\begin{equation}
\frac{D_{well}}{N}=-\frac{1}{V_{a}}\frac{d^{2}V}{ds^{2}}+...
\end{equation}
where $s = \Psi/\Psi_a=\rho^2$ is a normalized toroidal flux, $\Psi_a$ is the value of $\Psi$ at the plasma boundary, and $V_{a}$ is the volume of the entire plasma (not the $s$-dependent
volume). This way, $D_{Merc}/N$ will be $O\left(1\right)$ regardless
of the radius, pressure, or field and size scale of the equilibrium.
Solving for $N$,
\begin{equation}
N=-\frac{s_{\psi}\mu_{0}V_{a}}{\Psi_{a}^{3}}\frac{dp}{ds}\left(\int dS\frac{B^{2}}{\left|\nabla\Psi\right|^{3}}\right).
\end{equation}
Next, we can apply eq (4.11) from \cite{landreman2020magnetic}:
\begin{equation}
\int dS\frac{B^{2}}{\left|\nabla\Psi\right|^{3}}\approx\frac{\left|G\right|}{2\pi r^{2}B_{av}^{2}}
\end{equation}
where $G$ is the poloidal current outside the flux surface times $\mu_0/(2\pi)$, and $B_{av}$ is an average field strength.
Thus, we have
\begin{equation}
N=-\frac{\mu_{0}V_{a}\left|G\right|}{4\pi\left|\Psi_{a}\right|^{3}\rho r^{2}B_{av}^{2}}\frac{dp}{d\rho}.
\end{equation}
Using $r=a\rho$ where $a$ is defined in \cite{landreman2020magnetic} by $\left|\Psi_{a}\right|=\pi a^{2}B_{av}$,
the normalization is
\begin{equation}
N=-\frac{\mu_{0}V_{a}\left|G\right|}{4\left|\Psi_{a}\right|^{4}\rho^{3}B_{av}}\frac{dp}{d\rho}.
\end{equation}

In the optimizations that include a constraint on Mercier stability, the specific constraint we include is $\min_{s>0.2} (D_{Merc} / N) > 0.02$.
The minimum is taken over radii with $s > 0.2$ due to the inaccuracy of $D_{Merc}$ computed near the magnetic axis \cite{panici2023desc}.
The threshold of 0.02 is used as opposed to 0 to provide a small stability margin.

	
\bibliographystyle{iopart-num}

\bibliography{references.bib}

@article{kappel2024magnetic,
  title={The magnetic gradient scale length explains why certain plasmas require close external magnetic coils},
  author={Kappel, John and Landreman, Matt and Malhotra, Dhairya},
  journal={Plasma Physics and Controlled Fusion},
  volume={66},
  number={2},
  pages={025018},
  year={2024},
  publisher={IOP Publishing}
}

@article{glas2022global,
  title={Global stochastic optimization of stellarator coil configurations},
  author={Glas, Silke and Padidar, Misha and Kellison, Ariel and Bindel, David},
  journal={Journal of Plasma Physics},
  volume={88},
  number={2},
  pages={905880208},
  year={2022},
  publisher={Cambridge University Press}
}

@article{jorge2022single,
  title={A single-field-period quasi-isodynamic stellarator},
  author={Jorge, Rogerio and Plunk, GG and Drevlak, M and Landreman, M and Lobsien, J-F and Mata, K Camacho and Helander, P},
  journal={Journal of Plasma Physics},
  volume={88},
  number={5},
  pages={175880504},
  year={2022},
  publisher={Cambridge University Press}
}

@article{dudt2020desc,
  title={DESC: A stellarator equilibrium solver},
  author={Dudt, DW and Kolemen, E},
  journal={Physics of Plasmas},
  volume={27},
  number={10},
  year={2020},
  publisher={AIP Publishing}
}

@article{wiedman2024coil,
  title={Coil optimization for quasi-helically symmetric stellarator configurations},
  author={Wiedman, A and Buller, Stefan and Landreman, Matt},
  journal={Journal of Plasma Physics},
  volume={90},
  number={3},
  pages={905900307},
  year={2024},
  publisher={Cambridge University Press}
}

@article{ku2008physics,
  title={Physics design for ARIES-CS},
  author={Ku, Long-Poe and Garabedian, PR and Lyon, J and Turnbull, A and Grossman, A and Mau, TK and Zarnstorff, M and ARIES Team},
  journal={Fusion Science and Technology},
  volume={54},
  number={3},
  pages={673--693},
  year={2008},
  publisher={Taylor \& Francis}
}

@article{gori2001alpha,
  title={$\alpha$-particle confinement optimization in quasi-axisymmetric configurations},
  author={Gori, S and N{\"u}hrenberg, J and Zille, R and Okamura, S and Matsuoka, K and Murakami, S},
  journal={Plasma physics and controlled fusion},
  volume={43},
  number={2},
  pages={137--144},
  year={2001}
}

@article{goodman2024quasi,
  title={Quasi-isodynamic stellarators with low turbulence as fusion reactor candidates},
  author={Goodman, Alan G and Xanthopoulos, Pavlos and Plunk, Gabriel G and Smith, H{\aa}kan and N{\"u}hrenberg, Carolin and Beidler, Craig D and Henneberg, Sophia A and Roberg-Clark, Gareth and Drevlak, Michael and Helander, Per},
  journal={PRX Energy},
  volume={3},
  number={2},
  pages={023010},
  year={2024},
  publisher={APS}
}

@article{schilling2025numerics,
  title={The numerics of vmec++},
  author={Schilling, Jonathan},
  journal={arXiv preprint arXiv:2502.04374},
  year={2025}
}

@article{giuliani2024direct,
  title={Direct stellarator coil design using global optimization: application to a comprehensive exploration of quasi-axisymmetric devices},
  author={Giuliani, Andrew},
  journal={Journal of Plasma Physics},
  volume={90},
  number={3},
  pages={905900303},
  year={2024},
  publisher={Cambridge University Press}
}

@article{giuliani2024comprehensive,
  title={A comprehensive exploration of quasisymmetric stellarators and their coil sets},
  author={Giuliani, Andrew and Rodríguez, Eduardo and Spivak, Marina},
  journal={Journal of Plasma Physics},
  volume={91},
  DOI={10.1017/S0022377825000509},
  number={5},
  year={2025}, 
  pages={E128}
}

@article{scikit-learn,
  title={Scikit-learn: Machine Learning in {P}ython},
  author={Pedregosa, F. and Varoquaux, G. and Gramfort, A. and Michel, V.
          and Thirion, B. and Grisel, O. and Blondel, M. and Prettenhofer, P.
          and Weiss, R. and Dubourg, V. and Vanderplas, J. and Passos, A. and
          Cournapeau, D. and Brucher, M. and Perrot, M. and Duchesnay, E.},
  journal={Journal of Machine Learning Research},
  volume={12},
  pages={2825--2830},
  year={2011}
}

@article{zenodo,
author={M. Landreman},
journal = { \url{https://doi.org/10.5281/zenodo.20733436}},
year=2026
}

@article{landreman2022magnetic,
  title={Magnetic fields with precise quasisymmetry for plasma confinement},
  author={Landreman, Matt and Paul, Elizabeth},
  journal={Physical Review Letters},
  volume={128},
  number={3},
  pages={035001},
  year={2022},
  publisher={APS}
}

@article{landreman2025does,
  title={How does ion temperature gradient turbulence depend on magnetic geometry? Insights from data and machine learning},
  author={Landreman, Matt and Choi, Jong Youl and Alves, Caio and Balaprakash, Prasanna and Churchill, Michael and Conlin, Rory and Roberg-Clark, Gareth},
  journal={Journal of Plasma Physics},
  volume={91},
  number={4},
  year={2025}
}

@article{jang2026exponential,
  title={Exponential spectral scaling: robust and efficient stellarator boundary optimisation via mode-dependent scaling},
  author={Jang, Byoungchan and Landreman, Matt and Conlin, Rory},
  journal={Journal of Plasma Physics},
  volume={92},
  number={1},
  pages={E15},
  year={2026},
  publisher={Cambridge University Press}
}

@article{cadena2025constellaration,
  title={ConStellaration: A dataset of QI-like stellarator plasma boundaries and optimization benchmarks},
  author={Cadena, Santiago A and Merlo, Andrea and Laude, Emanuel and Bauer, Alexander and Agrawal, Atul and Pascu, Maria and Savtchouk, Marija and Guiraud, Enrico and Bonauer, Lukas and Hudson, Stuart and others},
  journal={Advances in Neural Information Processing Systems},
  volume={38},
  year={2026}
}

@article{sanchez2023quasi,
  title={A quasi-isodynamic configuration with good confinement of fast ions at low plasma $\beta$},
  author={S{\'a}nchez, E and Velasco, JL and Calvo, I and Mulas, S},
  journal={Nuclear Fusion},
  volume={63},
  number={6},
  pages={066037},
  year={2023},
  publisher={IOP Publishing}
}

@article{goodman2023constructing,
  title={Constructing precisely quasi-isodynamic magnetic fields},
  author={Goodman, Alan G and Mata, K Camacho and Henneberg, Sophia A and Jorge, Rogerio and Landreman, Matt and Plunk, GG and Smith, HM and Mackenbach, RJJ and Beidler, CD and Helander, P},
  journal={Journal of Plasma Physics},
  volume={89},
  number={5},
  pages={905890504},
  year={2023},
  publisher={Cambridge University Press}
}

@article{hirshman1985optimized,
  title={Optimized Fourier representations for three-dimensional magnetic surfaces},
  author={Hirshman, SP and Meier, HK},
  journal={The Physics of Fluids},
  volume={28},
  number={5},
  pages={1387--1391},
  year={1985},
  publisher={AIP Publishing}
}

@article{lion2025stellaris,
  title={Stellaris: A high-field quasi-isodynamic stellarator for a prototypical fusion power plant},
  author={Lion, J and Angl{\`e}s, J-C and Bonauer, L and Navarro, A Ba{\~n}{\'o}n and Ceron, SA Cadena and Davies, R and Drevlak, M and Foppiani, N and Geiger, J and Goodman, A and others},
  journal={Fusion Engineering and Design},
  volume={214},
  pages={114868},
  year={2025},
  publisher={Elsevier}
}

@article{hegna2025infinity,
  title={The infinity two fusion pilot plant baseline plasma physics design},
  author={Hegna, CC and Anderson, DT and Andrew, EC and Ayilaran, A and Bader, A and Bohm, TD and Mata, K Camacho and Canik, JM and Carbajal, L and Cerfon, A and others},
  journal={Journal of Plasma Physics},
  volume={91},
  number={3},
  pages={E76},
  year={2025},
  publisher={Cambridge University Press}
}

@article{schmitt2025magnetohydrodynamic,
  title={Magnetohydrodynamic equilibrium and stability properties of the Infinity Two fusion pilot plant},
  author={Schmitt, JC and Anderson, DT and Andrew, EC and Bader, A and Mata, K Camacho and Canik, JM and Carbajal, L and Cerfon, A and Cooper, WA and Davila, NM and others},
  journal={Journal of Plasma Physics},
  volume={91},
  number={3},
  pages={E88},
  year={2025},
  publisher={Cambridge University Press}
}

@article{landreman2026efficient,
  title={Efficient calculation of magnetic fields from ferromagnetic materials near strong electromagnets, and application to stellarator coil optimization},
  author={Landreman, Matt and Torreblanca, Humberto and Cerfon, Antoine},
  journal={Fusion Engineering and Design},
  volume={224},
  pages={115627},
  year={2026},
  publisher={Elsevier}
}

@article{law2022accelerating,
  title={Accelerating the estimation of collisionless energetic particle confinement statistics in stellarators using multifidelity Monte Carlo},
  author={Law, Frederick and Cerfon, Antoine and Peherstorfer, Benjamin},
  journal={Nuclear Fusion},
  volume={62},
  number={7},
  pages={076019},
  year={2022},
  publisher={IOP Publishing}
}

@article{alonso2022physics,
  title={Physics design point of high-field stellarator reactors},
  author={Alonso, JA and Calvo, I and Carralero, D and Velasco, JL and Garc{\'\i}a-Rega{\~n}a, JM and Palermo, I and Rapisarda, D},
  journal={Nuclear Fusion},
  volume={62},
  number={3},
  pages={036024},
  year={2022},
  publisher={IOP Publishing}
}

@article{guttenfelder2025predictions,
  title={Predictions of core plasma performance for the Infinity Two fusion pilot plant},
  author={Guttenfelder, W and Mandell, NR and Le Bars, G and Singh, L and Bader, A and Mata, K Camacho and Canik, JM and Carbajal, L and Cerfon, A and Davila, NM and others},
  journal={Journal of Plasma Physics},
  volume={91},
  number={3},
  pages={E83},
  year={2025},
  publisher={Cambridge University Press}
}

@article{bindel2023direct,
  title={Direct optimization of fast-ion confinement in stellarators},
  author={Bindel, David and Landreman, Matt and Padidar, Misha},
  journal={Plasma Physics and Controlled Fusion},
  volume={65},
  number={6},
  pages={065012},
  year={2023},
  publisher={IOP Publishing}
}

@article{anderson1994stellarator,
  title={A stellarator configuration for reactor studies},
  author={Anderson, D. T. and Garabedian, P. R.},
  journal={Nuclear Fusion},
  volume={34},
  number={6},
  pages={881--885},
  year={1994}
}

@article{garabedian1995reduction,
  title={Reduction of bootstrap current in the Modular Helias-like Heliac stellarator},
  author={Garabedian, PR and Gardner, HJ},
  journal={Physics of Plasmas},
  volume={2},
  number={6},
  pages={2020--2025},
  year={1995},
  publisher={American Institute of Physics}
}

@misc{czekanski2026catapult,
      title={{CATAPULT: A CUDA-Accelerated Timestepper for Alpha Particles Using Local Tricubics}}, 
      author={Michael Czekanski and Alexey R. Knyazev and David Bindel and Elizabeth J. Paul},
      year={2026},
      eprint={2604.07617},
      archivePrefix={arXiv},
      primaryClass={physics.comp-ph},
      url={https://arxiv.org/abs/2604.07617}, 
}

@inproceedings{olson2025ax,
  title = {{Ax: A Platform for Adaptive Experimentation}},
  author = {
    Olson, Miles and Santorella, Elizabeth and Tiao, Louis C. and
    Cakmak, Sait and Garrard, Mia and Daulton, Samuel and
    Lin, Zhiyuan Jerry  and Ament, Sebastian and Beckerman, Bernard and
    Onofrey, Eric and Igusti, Paschal and Lara, Cristian and
    Letham, Benjamin and Cardoso, Cesar and Shen, Shiyun Sunny and
    Lin, Andy Chenyuan and Grange, Matthew and Kashtelyan, Elena and
    Eriksson, David and Balandat, Maximilian and Bakshy, Eytan.
  },
  booktitle = {AutoML 2025 ABCD Track},
  year = {2025}
}

@article{eriksson2019scalable,
  title={Scalable global optimization via local Bayesian optimization},
  author={Eriksson, David and Pearce, Michael and Gardner, Jacob and Turner, Ryan D and Poloczek, Matthias},
  journal={Advances in neural information processing systems},
  volume={32},
  year={2019}
}

@inproceedings{hvarfner2024vanilla,
  title={Vanilla Bayesian optimization performs great in high dimensions},
  author={Hvarfner, Carl and Hellsten, Erik O and Nardi, Luigi},
  booktitle={Proceedings of the 41st International Conference on Machine Learning},
  pages={20793--20817},
  year={2024}
}

@article{velasco2024piecewise,
  title={Piecewise omnigenous stellarators},
  author={Velasco, JL and Calvo, I and Escoto, FJ and S{\'a}nchez, E and Thienpondt, H and Parra, FI},
  journal={Physical review letters},
  volume={133},
  number={18},
  pages={185101},
  year={2024},
  publisher={APS}
}

@article{liu2026optimization,
  title={Optimization of stellarator configurations combining omnigenity and piecewise omnigenity},
  author={Liu, Hengqian and Yu, Guodong and Velasco, Jos{\'e} Luis and Zhu, Caoxiang},
  journal={arXiv preprint arXiv:2603.12139},
  year={2026}
}

@article{velasco2026combination,
  title={Combination of quasi-isodynamic and piecewise omnigenous magnetic fields},
  author={Velasco, Jos{\'e} Luis and Calvo, Iv{\'a}n and Fern{\'a}ndez-Pacheco, V and Padidar, M and Liu, H and S{\'a}nchez, E and Yu, G and Zhu, C},
  journal={arXiv preprint arXiv:2603.12377},
  year={2026}
}

@article{landreman2020magnetic,
  title={Magnetic well and Mercier stability of stellarators near the magnetic axis},
  author={Landreman, Matt and Jorge, Rogerio},
  journal={Journal of Plasma Physics},
  volume={86},
  number={5},
  pages={905860510},
  year={2020},
  publisher={Cambridge University Press}
}

@article{swanson2025overview,
  title={Overview of the Helios Design: A Practical Planar Coil Stellarator Fusion Power Plant},
  author={Swanson, CPS and Kumar, STA and Dudt, DW and Flom, ER and Kalb, WB and Kruger, TG and Martin, MF and Olatunji, JR and Pasmann, S and Tang, LZ and others},
  journal={arXiv preprint arXiv:2512.08027},
  year={2025}
}

@article{najmabadi2008aries,
  title={The ARIES-CS compact stellarator fusion power plant},
  author={Najmabadi, F and Raffray, AR and Abdel-Khalik, SI and Bromberg, L and Crosatti, L and El-Guebaly, L and Garabedian, PR and Grossman, AA and Henderson, D and Ibrahim, A and others},
  journal={Fusion Science and Technology},
  volume={54},
  number={3},
  pages={655--672},
  year={2008},
  publisher={Taylor \& Francis}
}

@misc{ali2025poster,
  author       = {Ali, Issra and Goodman, Alan and Zingg, David},
  title        = {A {B}-spline parametrization for stellarator boundary optimization},
  howpublished = {Poster presented at the 67th {APS} Division of Plasma Physics Meeting},
  year         = {2025},
  address      = {Long Beach, {USA}}
}

@article{panici2023desc,
  title={The DESC stellarator code suite. Part 1. Quick and accurate equilibria computations},
  author={Panici, Dario and Conlin, Rory and Dudt, Daniel W and Unalmis, Kaya and Kolemen, Egemen},
  journal={Journal of Plasma Physics},
  volume={89},
  number={3},
  pages={955890303},
  year={2023},
  publisher={Cambridge University Press}
}

@article{dudt2024magnetic,
  title={Magnetic fields with general omnigenity},
  author={Dudt, Daniel W and Goodman, Alan G and Conlin, Rory and Panici, Dario and Kolemen, Egemen},
  journal={Journal of Plasma Physics},
  volume={90},
  number={1},
  pages={905900120},
  year={2024},
  publisher={Cambridge University Press}
}

@misc{liu2026optimizingstellaratorshiddensymmetry,
      title={Optimizing stellarators with hidden symmetry}, 
      author={Hengqian Liu and Guodong Yu and Caoxiang Zhu and José Luis Velasco and Rahul Gaur and Dario Panici and Egemen Kolemen and Mingyang Yu and Weixing Ding and Shaojie Wang and Ge Zhuang},
      year={2026},
      eprint={2502.09350},
      archivePrefix={arXiv},
      primaryClass={physics.plasm-ph},
      url={https://arxiv.org/abs/2502.09350}, 
}

@article{paul2022energetic,
  title={Energetic particle loss mechanisms in reactor-scale equilibria close to quasisymmetry},
  author={Paul, Elizabeth J and Bhattacharjee, A and Landreman, M and Alex, D and Velasco, JL and Nies, R},
  journal={Nuclear Fusion},
  volume={62},
  number={12},
  pages={126054},
  year={2022},
  publisher={IOP Publishing}
}

@article{bader2019stellarator,
  title={Stellarator equilibria with reactor relevant energetic particle losses},
  author={Bader, Aaron and Drevlak, M and Anderson, DT and Faber, BJ and Hegna, CC and Likin, KM and Schmitt, JC and Talmadge, JN},
  journal={Journal of Plasma Physics},
  volume={85},
  number={5},
  pages={905850508},
  year={2019},
  publisher={Cambridge University Press}
}

@article{bader2021modeling,
  title={Modeling of energetic particle transport in optimized stellarators},
  author={Bader, A and Anderson, DT and Drevlak, M and Faber, BJ and Hegna, CC and Henneberg, S and Landreman, M and Schmitt, JC and Suzuki, Y and Ware, A},
  journal={Nuclear Fusion},
  volume={61},
  number={11},
  pages={116060},
  year={2021},
  publisher={IOP Publishing}
}

@article{schuett2024exploring,
  title={Exploring novel compact quasi-axisymmetric stellarators},
  author={Schuett, Tobias M and Henneberg, Sophia A},
  journal={Physical Review Research},
  volume={6},
  number={4},
  pages={L042052},
  year={2024},
  publisher={APS}
}

@article{wei2026low,
  title={Low-dimensional geometry learning for turbulence prediction in optimized stellarators},
  author={Wei, Xishuo and Huang, Handi and Chen, Haotian and Zhu, Hongxuan and Bai, Zhe and Williams, Samuel and Lin, Zhihong},
  journal={arXiv preprint arXiv:2603.17366},
  year={2026}
}

@article{velasco2025exploration,
  title={Exploration of the parameter space of piecewise omnigenous stellarator magnetic fields},
  author={Velasco, JL and S{\'a}nchez, E and Calvo, I},
  journal={Nuclear Fusion},
  volume={65},
  number={5},
  pages={056012},
  year={2025},
  publisher={IOP Publishing}
}

@article{beidler2001stochastic,
  title={Stochastic diffusion of energetic ions in optimized stellarators},
  author={Beidler, CD and Kolesnichenko, Ya I and Marchenko, VS and Sidorenko, IN and Wobig, H},
  journal={Physics of Plasmas},
  volume={8},
  number={6},
  pages={2731--2738},
  year={2001},
  publisher={American Institute of Physics}
}

\end{document}